\documentclass[pdflatex,sn-mathphys-num]{sn-jnl}

\usepackage{graphicx}%
\usepackage{multirow}%
\usepackage{amsmath,amssymb,amsfonts}%
\usepackage{amsthm}%
\usepackage{mathrsfs}%
\usepackage[title]{appendix}%
\usepackage{xcolor}%
\usepackage{textcomp}%
\usepackage{manyfoot}%
\usepackage{booktabs}%
\usepackage{algorithm}%
\usepackage{algorithmicx}%
\usepackage{algpseudocode}%
\usepackage{listings}%
\usepackage{caption}%
\usepackage{lineno}
\theoremstyle{thmstyleone}%
\theoremstyle{thmstyletwo}%

\theoremstyle{thmstylethree}%

\usepackage{bibunits}

\usepackage{xr-hyper}
\begin{document}

\title[Article Title]{Re-emergence of fingering in solidifying liquid-metal drops}


\author[1,3]{\fnm{Siyu} \sur{Zhang}}
\equalcont{These authors contributed equally to this work.}

\author[2]{\fnm{Feng} \sur{Wang}}
\equalcont{These authors contributed equally to this work.}

\author[1]{\fnm{Yuan} \sur{Ma}}

\author[1]{\fnm{Hao} \sur{Jiang}}

\author*[1]{\fnm{Kai Leong} \sur{Chong}}\email{klchong@shu.edu.cn}

\author*[1]{\fnm{Quan} \sur{Zhou}}\email{qzhou@shu.edu.cn}

\author[2]{\fnm{Chao} \sur{Sun}}

\author*[3]{\fnm{Jianhua} \sur{Zhang}}\email{jhzhang@shu.edu.cn}

\affil[1]{\orgdiv{Shanghai Key Laboratory of Mechanics in Energy Engineering, Shanghai Institute of Applied Mathematics and Mechanics, School of Mechanics and Engineering Science}, \orgname{Shanghai University},  
\orgaddress{\city{Shanghai}, \postcode{200072},  \country{China}}}

\affil[2]{\orgdiv{New Cornerstone Science Laboratory, Center for Combustion Energy,
Department of Energy and Power Engineering}, \orgname{Tsinghua University}, \orgaddress{\city{Beijing}, \state{100084}, \country{China}}}

\affil[3]{\orgdiv{School of Microelectronics, Key Laboratory of Advanced Display and System Application of Ministry of Education}, \orgname{ Shanghai University}, \orgaddress{\city{Shanghai}, \state{200072}, \country{China}}}


\abstract{Solidification is expected to suppress fingering by removing mobile liquid. Here we report an unexpected re-emergence of fingering in impacting molten-tin drops: cooling first weakens peripheral fingers, but stronger undercooling restores them. At weak undercooling, fingers originate from a decelerating liquid rim and solidification reduces the effective acceleration that drives Rayleigh--Taylor-like amplification. At strong undercooling, solidification structures develop while liquid is still spreading, creating lateral mobility contrasts that redirect the residual flow. Comparison between the kinematically estimated solidification time and the experimentally measured fingering-onset and growth-termination times reveals a reversal in temporal ordering across the temperature range. A physics-based reduced-order model combines solidification-modified rim deceleration with mobility-gradient growth, captures the characteristic onset dynamics and accounts for the final finger length within the same physical framework. These findings establish solidification as an active means of regulating interfacial flow and morphology, with broad implications spanning metal-based additive manufacturing and debris mitigation in semiconducting industry.}


\maketitle
Drop impact redistributes liquid over millisecond timescales and is central to processes ranging from spray coating and thermal-spray deposition to droplet-based additive manufacturing, and it is particularly important when the droplet becomes liquid metal which have broad relevance in semiconducting industry \cite{lohse2022fundamental,debroy2019metalprinting,chong2026euv,chandra2009formation}. In all of these settings, the final outcome depends not only on how far the drop spreads, but also on how the contact region, rim and internal flow evolve before motion is arrested or the liquid solidifies \cite{yarin2006drop,josserand2016drop,sanjay2025unifying}. Understanding drop impact therefore requires resolving the transient interplay between inertia, capillarity and interfacial conditions that selects the final morphology.

Phase change introduces an additional layer of complexity because the interfacial boundary condition can evolve on the same timescale as the impact itself. On sufficiently hot substrates, evaporation and vapour-film formation can substantially modify liquid--solid contact and give rise to dynamic Leidenfrost behaviour \cite{quere2013leidenfrost,tran2012superheated,shirota2016dynamic,jiang2022inhibiting}. On cold substrates, by contrast, solidification can proceed while the lamella is still spreading. Previous studies have shown that increasing cooling suppresses spreading, promotes contact-line arrest and preserves an earlier liquid morphology \cite{huerre2025freezing,deruiter2017contact,kant2020fastfreezing,gielen2020solidification,grivet2022contact,lolla2022arrested}. These observations support an intuitive picture in which solidification primarily stabilizes the impact by progressively removing mobile liquid.

This stabilizing picture, however, treats the newly formed solid mainly as liquid that has ceased to move, rather than as an evolving boundary for the liquid that remains mobile. Spreading drops are known to be highly sensitive to the structure of the underlying boundary: engineered textures can alter wetting and contact dynamics, while spatial variations in confinement can redirect flow and control interfacial fingering \cite{bird2013reducing,quere2008wetting,alhousseiny2012geometry,zheng2015control,PhysRevLett.125.184501}. During solidifying impact, such a structured boundary can arise spontaneously. If ridges or other non-uniform solid structures develop before spreading ceases, they can locally modify the thickness and hydraulic mobility of the remaining liquid, thereby redistributing the radial flow. This raises a distinct question: can solidification-generated structures feed back on the spreading liquid strongly enough to reverse the otherwise stabilizing effect of freezing?

Here we show that molten-tin drops impacting cooled steel exhibit an unexpected non-monotonic response: fingering is first suppressed as the substrate is cooled, but re-emerges under stronger cooling. Time-resolved imaging reveals two distinct dynamical sequences. At weak undercooling, a decelerating liquid rim destabilizes before solidification substantially restructures the flow. At strong undercooling, by contrast, ridge-like solidification structures develop beneath a still-moving lamella before persistent fingers emerge. We show that these observations arise from two competing consequences of solidification: a reduction of the effective Rayleigh--Taylor (RT) driving in the weak-undercooling regime, and the generation of lateral mobility contrasts in the strong-undercooling regime. A physics-based reduced-order model captures the resulting onset dynamics and predicts the final finger length, providing a compact physical framework for understanding morphology selection during rapid solidification.

\section{Cooling reverses the fingering response}
\begin{figure*}
  \centering
  \includegraphics[width=1\textwidth]{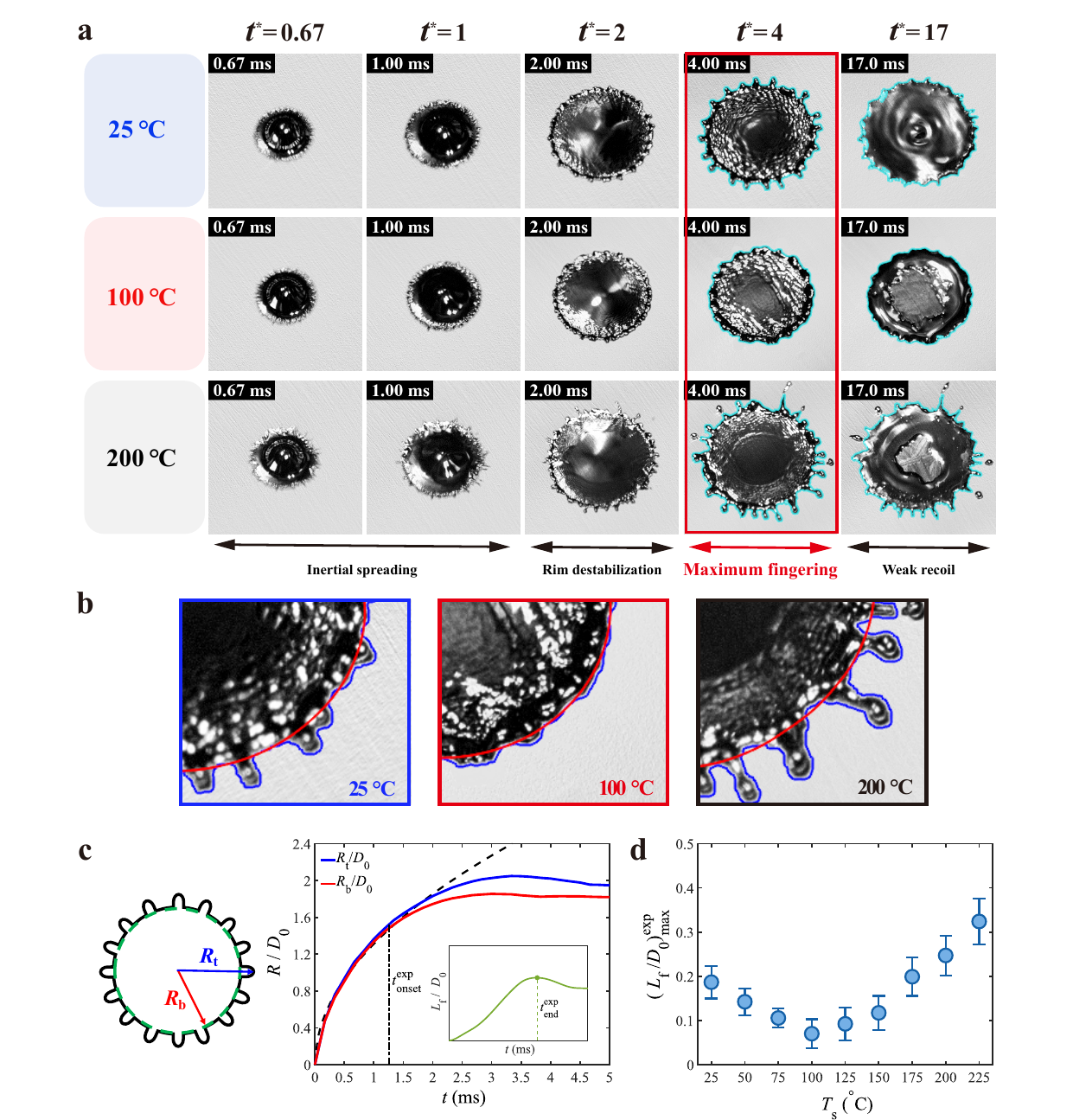}
  \caption{\textbf{Cooling suppresses and then restores fingering.} \textbf{a}, Representative plan-view sequences for $T_{\mathrm{s}}=25$, $100$ and $200\,^{\circ}\mathrm{C}$. Early inertial spreading is similar, whereas the later morphology differs markedly: RT-like rim corrugations are suppressed at intermediate cooling and renewed fingering is preceded by irregular solidification structures at strong cooling. \textbf{b,} Magnified views of the droplet periphery at $t^*\simeq4$. The blue and red curves denote the overall and main-body contours, respectively. \textbf{c}, Definition and representative evolution of $R_{\mathrm b}$, $R_{\mathrm t}$, $t_{\mathrm{onset}}^{\mathrm{exp}}$ and $t_{\mathrm{end}}^{\mathrm{exp}}$. \textbf{d}, Non-monotonic maximum normalized finger length versus substrate temperature. Symbols and bars denote the mean and variability based on eight to ten repeated experiments}.
  \label{fig:1}
\end{figure*}

Tin drops with initial diameter $D_0=3.0\pm0.05\,\mathrm{mm}$ impact polished steel at $V_0\simeq3.0\,\mathrm{m\,s^{-1}}$ (see Methods). The corresponding inertial time is $t_{\mathrm{i}}=D_0/V_0\simeq1.0\,\mathrm{ms}$, and we use the normalized time $t^*=t/t_{\mathrm{i}}$. The chamber pressure is maintained at $60$--$70\,\mathrm{Pa}$ for all experiments, so changing the substrate temperature $T_{\mathrm{s}}$ mainly changes the solidification condition rather than gas-mediated impact dynamics \cite{xu2005splashing,mandre2009precursors,driscoll2011air,sprittles2024gas}.

Fig.~\ref{fig:1}a compares impacts at $T_{\mathrm{s}}=25$, $100$ and $200\,^{\circ}\mathrm{C}$. Up to $t\sim t_{\mathrm{i}}$, all three drops flatten rapidly and form comparable lamellae and peripheral rims. The temperature dependence becomes clear only afterwards. At $200\,^{\circ}\mathrm{C}$, azimuthal rim corrugations are already visible by $t^*\simeq2$, and long fingers are clearly established by $t^*=4$, consistent with the deceleration-driven instability reported for impacting molten metals \cite{mehdizadeh2004formation,gielen2020solidification,lherm2021rayleigh}. Notably, some fingers undergo pronounced necking and incipient breakup at $200\,^\circ\mathrm{C}$, making them appear considerably longer in Fig.~\ref{fig:1}a, but these breakup-associated fingers are excluded from the mean finger-length statistics. Cooling to $100\,^{\circ}\mathrm{C}$ produces the expected stabilizing effect: the early lamella remains similar, but the corrugations stay weak and the final perimeter contains few pronounced fingers.

With further cooling to $25\,^{\circ}\mathrm{C}$, the response reverses. The rim at $t^*\simeq2$ is not yet markedly different from that at $100\,^{\circ}\mathrm{C}$, but the interior of the lamella is already more irregular. By $t^*\simeq4$, branched and non-axisymmetric solidification structures are visible while the film is still spreading, and pronounced fingers subsequently re-emerge at the perimeter. At the two larger temperature cases, the central frozen feature is instead approximately circular, consistent with capillary-wave focusing and locally enhanced thinning \cite{gielen2020solidification}. The magnified edge views at $t^*\simeq4$ further highlight this non-monotonic response in Fig.~\ref{fig:1}b: pronounced protrusions are observed at both 25 and $200\,^\circ\mathrm{C}$, whereas the perimeter remains comparatively smooth at $100\,^\circ\mathrm{C}$.

We quantify the perimeter using the mean main-body radius $R_{\mathrm b}$, the mean finger-tip radius $R_{\mathrm t}$ (see Fig.~\ref{fig:1}c). During the initial inertial stage, $R_{\mathrm b}$ and $R_{\mathrm t}$ follow one another,  although small non-zero differences can arise from
pixel-scale edge fluctuations and uncertainty in the reconstruction of the main-body contour. Their later separation measures the growth of persistent protrusions. The experimentally measured fingering-onset time $t_{\mathrm{onset}}^{\mathrm{exp}}$ is defined as the first persistent time at which $R_{\mathrm t}-R_{\mathrm b}$ reaches the detection length
$L_{\mathrm{c}}=0.04D_0$. 
For the subsequent analysis, the reported finger length $L_{\mathrm{f}}$ is referenced to this onset state, such that $L_{\mathrm{f}}(t_{\mathrm{onset}}^{\exp})=0$. The experimental growth-termination time $t_{\mathrm{end}}^{\mathrm{exp}}$ is defined as the time at which the relative finger length $L_{\mathrm{f}}$ reaches its maximum, as illustrated in the inset of Fig.~\ref{fig:1}c.
Thresholds of $0.03D_0$ and $0.05D_0$ shift the absolute onset but preserve the temperature dependence (see Supplementary Section~\hyperref[onset_sensitivity]{SI~5}).

The maximum finger length reveals the central result in Fig.~\ref{fig:1}d. To assess the robustness of this trend, eight to ten independent impacts are performed at each substrate temperature. Cooling from $225$ to approximately $100\,^{\circ}\mathrm{C}$ progressively decreases $(L_{\mathrm{f}}/D_0)_{\max}$, whereas stronger cooling increases it again. Solidification therefore does not simply arrest the rim. 

\section{How solidification affects fingering pattern}

\begin{figure*}
  \centering
  \includegraphics[width=1\linewidth]{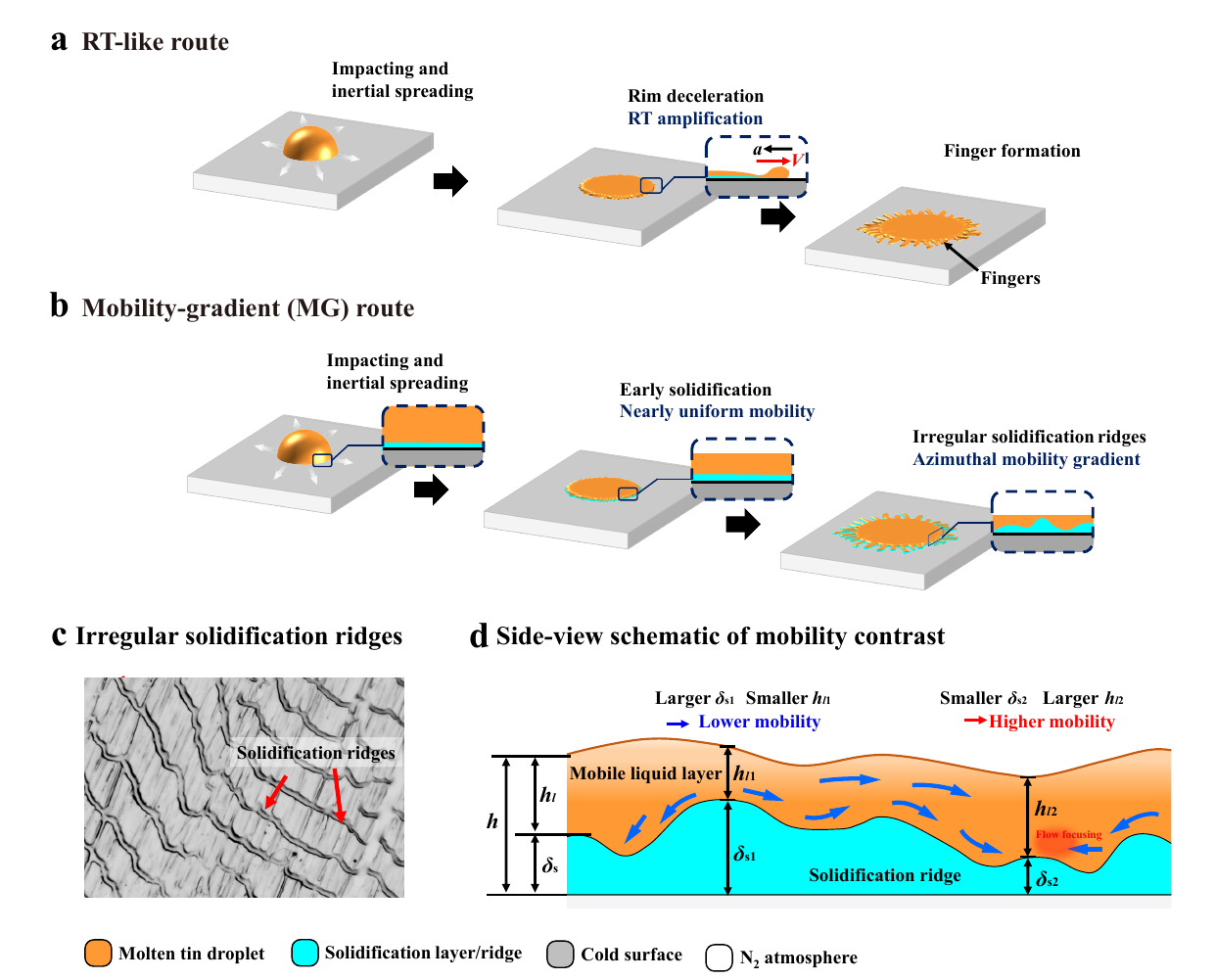}
  \caption{\textbf{Two routes to fingering during solidification.} \textbf{a}, RT-like route: deceleration of the liquid rim amplifies azimuthal disturbances; freezing weakens this amplification in the model. \textbf{b}, MG route: an uneven solid layer creates variations in liquid thickness and flow resistance, directing liquid towards preferential paths. \textbf{c}, The micrograph shows irregular annular solidification ridges on the frozen underside. \textbf{d}, Schematic interpretation of the mobility contrast. A locally thicker solidification region reduces the remaining mobile-liquid thickness, $h_{\mathrm{l}}=h-\delta_{\mathrm{s}}$, and hence the hydraulic mobility $M\propto h_{\mathrm{l}}^3/\mu_{\mathrm{l}}$, whereas a thinner solidification region retains a larger $h_{\mathrm{l}}$ and higher mobility, favouring preferential redistribution of the remaining liquid.}
  \label{fig:2}
\end{figure*}

Here we explain the non-monotonic response of fingering instability upon the changing undercooling. 
For a partially frozen lamella, the mobile-liquid thickness is $h_{\mathrm{l}}=h-\delta_{\mathrm{s}}$, which is the difference between the total lamella thickness $h$ and solid-layer thickness  $\delta_{\mathrm{s}}$, as illustrated in Fig.~\ref{fig:2}d. Lubrication theory gives the depth-integrated flux $\boldsymbol q\sim-(h_{\mathrm{l}}^3/\mu_{\mathrm{l}})\boldsymbol\nabla p$ and hence the hydraulic mobility $M\propto h_{\mathrm{l}}^3/\mu_{\mathrm{l}}$ \cite{oron1997thinfilms}, where $p$ is pressure and $\mu_{\mathrm{l}}$ is liquid viscosity. From the expression, one can see that freezing lowers mobility of the liquid layer. Besides, thickness variation can lead to mobility lateral contrast.

At weak undercooling, the solid layer stays thin and smooth, and fingers arise mainly from the RT-like instability of the decelerating rim (Fig.~\ref{fig:2}a) \cite{mehdizadeh2004formation,gielen2020solidification,lherm2021rayleigh}. In our model, freezing reduces effective rim deceleration and disturbance growth, so cooling suppresses this route.

At strong undercooling, ridges form beneath the spreading lamella (see Fig.~\ref{fig:2}c), potentially through crystal growth and stick--slip contact-line motion \cite{deruiter2017contact,gielen2020solidification,koldeweij2021initial}. The thinner liquid above a ridge offers greater resistance, redirecting flow towards regions with lower-resistance that feed local protrusions in Fig.~\ref{fig:2}b. We call this the mobility-gradient (MG) route.

The MG route requires solid structures to develop while the liquid remains mobile, making the solidification penetration ratio an important indicator of whether this condition is met. Assuming semi-infinite tin and steel bodies with negligible contact resistance, the initial interfacial temperature is \cite{gielen2020solidification,deruiter2018selfpeeling}

\begin{equation}
T_{\mathrm{i}}=\frac{e_{\mathrm{l}}T_{\mathrm{l}}+e_{\mathrm{sub}}T_{\mathrm{s}}}{e_{\mathrm{l}}+e_{\mathrm{sub}}},
\end{equation}
where $T_{\mathrm{l}}$ is the initial temperature of the liquid tin and $e_j=(K_j\rho_jc_{p,j})^{1/2}$ is the thermal effusivity of liquid tin ($j={\mathrm{l}}$) or substrate ($j=\mathrm{sub}$); $K_j$, $\rho_j$ and $c_{p,j}$ are thermal conductivity, density and specific heat capacity. The initial liquid-tin temperature is approximated by the equilibrium melting temperature of tin, $T_{\mathrm{l}}\simeq T_{\mathrm{m}} =232~^\circ\mathrm{C}$, consistent with the experimental procedure in which the molten tin is discharged shortly after melting. As $T_{\mathrm{s}}$ rises from $25$ to $225~\,^{\circ}\mathrm{C}$, $\Delta T_{\mathrm{i}}=T_{\mathrm{m}}-T_{\mathrm{i}}$ falls from $109.5$ to $3.7~\,\mathrm{K}$, and ${\mathrm{Ste}}_{\mathrm{s}}=c_{p,\mathrm{s}}\Delta T_{\mathrm{i}}/L_{\mathrm{m}}$ falls from $0.42$ to $0.014$. Here, ${\mathrm{Ste}}_{\mathrm{s}}$ measures the sensible-heat change associated with the interfacial undercooling relative to the latent heat of fusion, and therefore characterizes the strength of solidification undercooling.

At $t=t_{\mathrm{i}}$, the measured diameter $D_{\mathrm{i}}\approx2.5D_0$ gives a mean lamella thickness $\overline h_{\mathrm{i}}=2D_0/[3(D_{\mathrm{i}}/D_0)^2]\approx0.1D_0$ by volume conservation. A one-dimensional Stefan estimate of the solid-layer thickness is
\begin{equation}
\delta_{\mathrm{s}}(t)=\left(2\alpha_{\mathrm{s}}{\mathrm{Ste}}_{\mathrm{s}}t\right)^{1/2},
\end{equation}
where $\alpha_{\mathrm{s}}$ is the thermal diffusivity of solid tin.
The penetration ratio $\Phi_{\mathrm{i}}=\delta_{\mathrm{s}}(t_{\mathrm{i}})/\overline h_{\mathrm{i}}$ measures the solid-layer thickness relative to the mean lamella thickness. Defining the critical Stefan number as
$\mathrm{Ste}_{\mathrm{c}}=\overline h_{\mathrm{i}}^{2}/(2\alpha_{\mathrm{s}}t_{\mathrm{i}})$,
we obtain $\Phi_{\mathrm{i}}=(\mathrm{Ste}_{\mathrm{s}}/\mathrm{Ste}_{\mathrm{c}})^{1/2}$.
Substituting the present values of $h_{\mathrm{i}}$, $\alpha_{\mathrm{s}}$, and $t_{\mathrm{i}}$ gives $\mathrm{Ste}_{\mathrm{c}}\approx1$. At the coldest substrate temperature, $\Phi_{\mathrm{i}}\approx0.6<1$. Thus, within one inertial time, a substantial solid layer can coexist with a still-mobile liquid layer, providing the conditions for solidification to redirect the flow.

\section{Fingering onset reveals a crossover between two routes}

\begin{figure*} 
\centering
\includegraphics[width=1\linewidth]{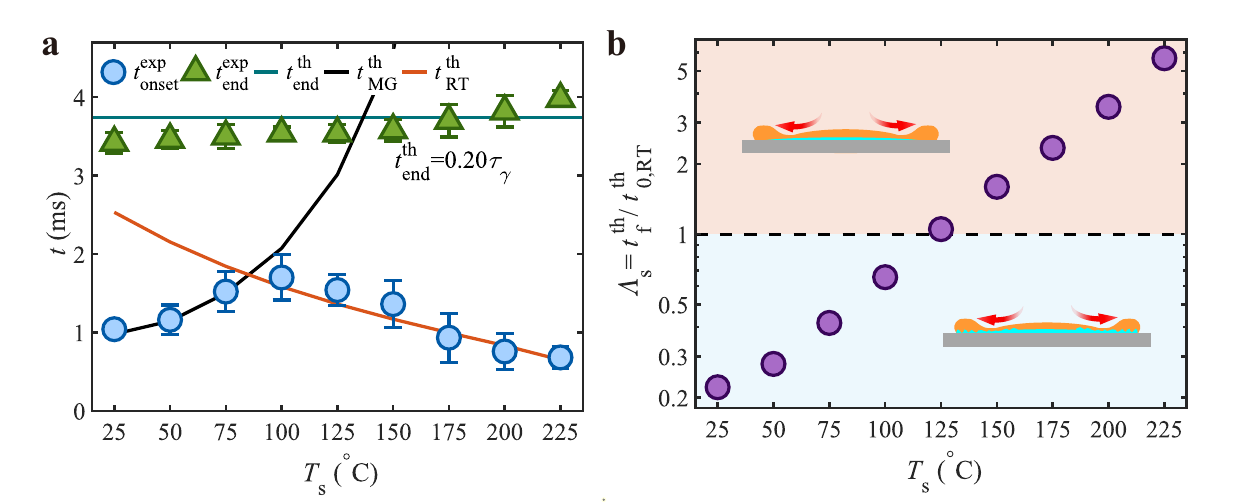}
\caption{
\textbf{Timing of fingering and solidification.}
\textbf{a}, Experimentally measured fingering-onset time
$t_{\mathrm{onset}}^{\mathrm{exp}}$ (blue circles) and growth-termination time
$t_{\mathrm{end}}^{\mathrm{exp}}$ (green triangles) as functions of substrate temperature
$T_{\mathrm{s}}$. Error bars indicate experimental variability.
The orange and black curves show the theoretical onset predictions for the RT-like
and MG routes from Eqs.~\eqref{eq:tRT} and~\eqref{eq:tMG}, respectively.
The horizontal line denotes the theoretical growth-termination time,
$t_{\mathrm{end}}^{\mathrm{th}}=0.20\tau_{\gamma}$.
\textbf{b}, Theoretically evaluated timing ratio
$\Lambda_{\mathrm{s}}
=t_{\mathrm{f}}^{\mathrm{th}}/t_{0,\mathrm{RT}}^{\mathrm{th}}$, quantifying the competition between the two routes. $\Lambda_{\mathrm{s}}<1$ (blue shading) indicates earlier initiation of the MG route,
whereas $\Lambda_{\mathrm{s}}>1$ (orange shading) indicates earlier initiation of the RT-like route.
The dashed line at $\Lambda_{\mathrm{s}}=1$ marks the transition between the two routes,
and the schematics illustrate the corresponding interfacial states.}
\label{fig:3}
\end{figure*}

To understand the non-monotonic fingering response, we examine the characteristic times associated with the experimentally observed onset and the development of the two proposed routes.
Fig.~\ref{fig:3}a reveals two distinct routes in the experimentally measured fingering-onset time $t_{\mathrm{onset}}^{\mathrm{exp}}$. As the substrate temperature increases from $25~^{\circ}\mathrm{C}$ to approximately $100~^{\circ}\mathrm{C}$, fingering is progressively delayed. At higher temperatures, however, the trend reverses and the onset advances with increasing $T_{\mathrm{s}}$. 
In contrast, the growth-termination times $t_{\mathrm{end}}^{\mathrm{exp}}$ exhibits only a weak dependence on $T_{\mathrm{s}}$. Compared with the pronounced variation in $t_{\rm onset}^{\rm exp}$, $t_{\rm end}^{\rm exp}$ decreases only slightly as $T_{\mathrm{s}}$ is reduced, and remains clustered around $t_{\mathrm{end}}^{\mathrm{th}}\simeq0.20\tau_\gamma\simeq3.74\,\mathrm{ms}$, where $\tau_\gamma=(\rho_{\mathrm{l}}D_0^3/\gamma)^{1/2}$ is the inertio-capillary time scale, representing the characteristic time over which inertia and surface tension govern droplet deformation \cite{zhang2022impact}. The temperature dependence of the fingering dynamics therefore arises primarily from the onset rather than the termination of finger growth.

We first model the fingering onset based on the conventional RT-like instability of the decelerating rim. Because image sequences allow one to obtain the onset time $t_{\mathrm{onset}}^{\mathrm{exp}}$ at which $R_{\mathrm t}-R_{\mathrm b}= L_{\mathrm{c}}$, the $t_{\mathrm{onset}}^{\mathrm{exp}}$ must include both the development of the instability and the subsequent growth required for the protrusion to reach $L_{\mathrm{c}}$.

For the RT-like route, rim deceleration amplifies disturbances while surface tension suppresses short wavelengths. The dispersion relation $\sigma_{\mathrm{RT}}^2(k)\sim a_{\mathrm{eff}}k-\gamma k^3/\rho_{\mathrm{l}}$ describes the growth of interfacial disturbances, where $\sigma_{\mathrm{RT}}(k)$ is the temporal growth rate of a disturbance with wavenumber $k$, and $a_{\mathrm{eff}}$ is the effective deceleration magnitude. The growth rate of  the fastest-growing mode as \cite{SHARP19843}
\begin{equation}
\sigma_{\mathrm{RT,max}}\sim \left(\frac{\rho_{\mathrm{l}} a_{\mathrm{eff}}^3}{\gamma}\right)^{1/4}.
\end{equation}

We represent freezing by an immobile lower layer and approximate the effective depth-averaged velocity as $V_{\mathrm{RT}}\sim V_0(1-\Phi_{\mathrm{i}})\simeq V_0(1-\sqrt{\mathrm{Ste}_{\mathrm{s}}})$, since $\mathrm{Ste}_{\mathrm{c}}\approx1$. Here, $V_0$ is the droplet impact velocity and is used as the characteristic velocity scale for the inertia-dominated spreading stage~\cite{mehdizadeh2004formation,aziz2000impact}. As a reduced-order approximation, we set $a_{\mathrm{eff}}\sim V_{\mathrm{RT}}^2/R_0$, where $R_0$ is the initial droplet radius \cite{aziz2000impact,gielen2020solidification}. This gives
\begin{equation}
\sigma_{\mathrm{RT,max}}\sim \sigma_{\mathrm{RT},0}(1-\sqrt{\mathrm{Ste}_{\mathrm{s}}})^{3/2},
\end{equation}
where $\sigma_{\mathrm{RT},0}$ is the fully liquid growth rate. Freezing therefore weakens RT-like amplification in this model.

The $\sigma_{\mathrm{RT, max} }$ obtained above provides the characteristic timescale over which the RT instability develops, and we use this to estimate the time required for the rim instability to become established. After that, the protrusion must continue to grow until it reaches the experimental detection length $L_{\mathrm{c}}$. Because $L_{\mathrm{c}}$ is small, the radial velocity during this short stage is approximated as constant, giving a growth time of ${L_{\mathrm{c}}/V_{\mathrm{RT}}}$. The theoretical fingering-onset time in RT route is therefore
\begin{equation}
t_{\mathrm{RT}}^{\mathrm{th}}=\frac{R_0/V_0}{(1-\sqrt{\mathrm{Ste}_{\mathrm{s}}})^{3/2}}+\frac{L_{\mathrm{c}}}{V_0(1-\sqrt{\mathrm{Ste}_{\mathrm{s}}})}.
\label{eq:tRT}
\end{equation}
The first term represents the initiation of RT amplification, whereas the second is the additional delay required for the protrusion to reach the experimental detection length. 

As shown by the orange line in Fig.~\ref{fig:3}a, the RT model reproduces the measured onset at weak undercooling. However, below approximately $100~^\circ\mathrm{C}$, the predicted $t_{\mathrm{RT}}^{\mathrm{th}}$ increases with decreasing $T_{\mathrm{s}}$, while the measured $t_{\mathrm{onset}}^{\mathrm{exp}}$ decreases. This opposite temperature dependence shows that the RT mechanism alone cannot explain the low-temperature branch, indicating that another process becomes important at stronger undercooling.

To determine when solidification becomes dynamically capable of modifying the spreading, we define the characteristic solidification time $t_{\mathrm{f}}^{\mathrm{th}}$ as the time when the spreading velocity $V_{\mathrm{b}}$ becomes comparable to the solidification-front velocity $U_{\mathrm{f}}=\kappa\Delta T_{\mathrm{i}}$, (see Section~\hyperref[solidification_time]{SI~4} for the details of the determination of $t_{\mathrm{f}}^{\mathrm{th}}$), 
where $\kappa$ is the kinetic coefficient of the solidification front, taken as $0.034~\mathrm{m\,s^{-1}\,K^{-1}}$\cite{gielen2020solidification}.

Once solidification becomes dynamically relevant, the resulting mobility variations can promote the growth of edge perturbations. To estimate this growth, we adopt the viscous--capillary scaling $\sigma_{\mathrm{v}}(k)\sim U_{\mathrm{f}}k-\gamma h_{\mathrm{l}}^2k^3/\mu_{\mathrm{l}}$, using $U_{\mathrm{f}}$ as the flow-speed scale at $t_{\mathrm{f}}^{\mathrm{th}}$ \cite{chuoke1959instability,homsy1987viscous}. With ${\mathrm{Ca}}_{\mathrm{f}}=\mu_{\mathrm{l}}U_{\mathrm{f}}/\gamma$, this gives $\sigma_{\mathrm {v,max}}\sim U_{\mathrm{f}}\sqrt{\mathrm{Ca}_{\mathrm{f}}}/h_{\mathrm{l}}$, which captures viscous–capillary growth but not the mobility variation induced by solidification. We therefore use the penetration ratio $\Phi_{\mathrm{i}}$ as a measure of the solidification effect. In the fully liquid limit, $\Phi_{\mathrm{i}} \rightarrow 0$, this contribution vanishes, whereas it increases with increasing $\Phi_{\mathrm{i}}$. The MG growth rate is therefore estimated as
\begin{equation}
\sigma_{\mathrm{MG,max}}\sim \Phi_{\mathrm{i}}\sigma_{\mathrm {v,max}},
\sim \frac{U_{\mathrm{f}}}{h_{\mathrm{l}}}\sqrt{\mathrm{Ca}_{\mathrm{f}}}\Phi_{{i}}
\sim \frac{U_{\mathrm{f}}}{h_{\mathrm{l}}}\sqrt{{\mathrm{Ca}}_{\mathrm{f}}\mathrm{Ste}_{\mathrm{s}}},
\end{equation}
where the last relation uses $\Phi_{\mathrm{i}} \simeq \sqrt{\mathrm{Ste}_{\mathrm{s}}}$ for the present conditions. Taking $h_{\mathrm{l}}$ as the length scale converts this rate into an estimated extension speed, $u_{\mathrm{MG}}\sim h_{\mathrm{l}}\sigma_{\mathrm{MG,max}}\sim U_{\mathrm{f}}\sqrt{\mathrm{Ca}_{\mathrm{f}}\mathrm{Ste}_{\mathrm{s}}}$. After the MG route becomes active at $t_{\mathrm{f}}^{\mathrm{th}}$, the protrusion must further grow by $L_{\mathrm{c}}$ at the characteristic speed $u_{\mathrm{MG}}$, giving

\begin{equation}
t_{\mathrm{MG}}^{\mathrm{th}}=t_{\mathrm{f}}^{\mathrm{th}}+\frac{L_{\mathrm{c}}}{u_{\mathrm{MG}}}.
\label{eq:tMG}
\end{equation}
Stronger undercooling advances $t_{\mathrm{f}}^{\mathrm{th}}$ and increases $u_{\mathrm{MG}}$, bringing detectable fingering forward. As shown by the black line in Fig.~\ref{fig:3}a, $t_{\mathrm{MG}}^{\mathrm{th}}$ decreases as $T_{\mathrm{s}}$ is lowered below approximately $100\,^\circ\mathrm{C}$, consistent with the measured $t_{\mathrm{onset}}^{\mathrm{exp}}$. The MG prediction therefore captures the low-temperature branch that cannot be explained by the RT-like model.

The theoretical fingering-onset time is therefore expressed as
\begin{equation}
t_{\mathrm{onset}}^{\mathrm{th}}
=
\begin{cases}
t_{\mathrm{RT}}^{\mathrm{th}}, & \text{RT-like route},\\[4pt]
t_{\mathrm{MG}}^{\mathrm{th}}, & \text{MG route}.
\end{cases}
\label{eq:main_tonset}
\end{equation}
Taken together, the RT-like prediction accounts for the weak-undercooling regime, whereas the MG correction accounts for the renewed advance of fingering at strong undercooling. The two routes therefore reproduce the non-monotonic variation of the fingering-onset time in Fig.~\ref{fig:3}a. The detailed estimates are given in Section~\hyperref[secSI6]{SI~6}.

The onset times $t_{\mathrm{RT}}^{\mathrm{th}}$ and $t_{\mathrm{MG}}^{\mathrm{th}}$ above are constructed to match the experimental detection criterion and therefore include the finite time required for a protrusion to grow to $L_{\mathrm{c}}$. To isolate the underlying initiation of the two routes, we remove this growth-to-detection delay, giving

\begin{equation}
t_{\mathrm{0,onset}}^{\mathrm{th}}
=
\left\{
\begin{aligned}
t_{\mathrm{0,RT}}^{\mathrm{th}}
&=
\frac{R_0/V_0}
{\left(1-\sqrt{\mathrm{Ste}_{\mathrm{s}}}\right)^{3/2}},
& \quad &\text{RT-like route},\\[6pt]
t_{\mathrm{0,MG}}^{\mathrm{th}}
&=
t_{\mathrm{f}}^{\mathrm{th}},
& \quad &\text{MG route},
\end{aligned}
\right.
\label{eq:threshold_free_times}
\end{equation}
where $t_{0,\mathrm{RT}}^{\mathrm{th}}$ and $t_{0,\mathrm{MG}}^{\mathrm{th}}$ represent the predicted times at which the RT-like and MG routes begin to amplify perturbations, before they reach the experimental detection length. To compare the competition between the two physical routes, we define
\begin{equation}
{\Lambda_{\mathrm{s}}}= t_{\mathrm{f}}^{\mathrm{th}}/t_{\mathrm{0,RT}}^{\mathrm{th}}.
\label{eq:ratio}
\end{equation}
The resulting $\Lambda_{\mathrm{s}}$ is shown in Fig.~\ref{fig:3}b. For $\Lambda_{\mathrm{s}}<1$, the MG route is initiated earlier; for $\Lambda_{\mathrm{s}}>1$, the RT-like route begins first. The crossing of $\Lambda_{\mathrm{s}}=1$ marks the transition between the MG and RT-like routes.

\section{Growth speed and duration set the finger length}

The final finger length depends on how fast the finger extends and how long it can grow. The reported $L_{\mathrm{f}}$ measures extension beyond the detection length $L_{\mathrm{c}}$, so its maximum is the additional growth between $t_{\mathrm{onset}}^{\mathrm{th}}$ and $t_{\mathrm{end}}^{\mathrm{th}}$. Approximating the relative extension speed by a constant $u_{\mathrm{sel}}$ gives, when onset precedes growth termination,

\begin{equation}
\left(\frac{L_{\mathrm{f}}}{D_0}\right)_{\max}^{\mathrm{th}}
=\frac{u_{\mathrm{sel}}}{D_0}
\left(t_{\mathrm{end}}^{\mathrm{th}}-t_{\mathrm{onset}}^{\mathrm{th}}\right).
\label{eq:Lmax}
\end{equation}

We assume that the route setting onset also controls subsequent growth. For the RT-like route, we use $u_{\mathrm{sel}}=\eta_{\mathrm{RT}}V_0(1-\sqrt{\mathrm{Ste}_{\mathrm{s}}})$. The factor $\eta_{\mathrm{RT}}=0.1$, obtained from the measured RT-like extension speeds, converts the radial transport scale into finger-tip motion relative to the main-body edge. For the MG route, we use the estimated protrusion speed $u_{\mathrm{sel}}=u_{\mathrm{MG}}$ (see Section~\hyperref[max_finger_length_model]{SI~7}).

\begin{figure*} 
  \centering
  \includegraphics[width=1\linewidth]{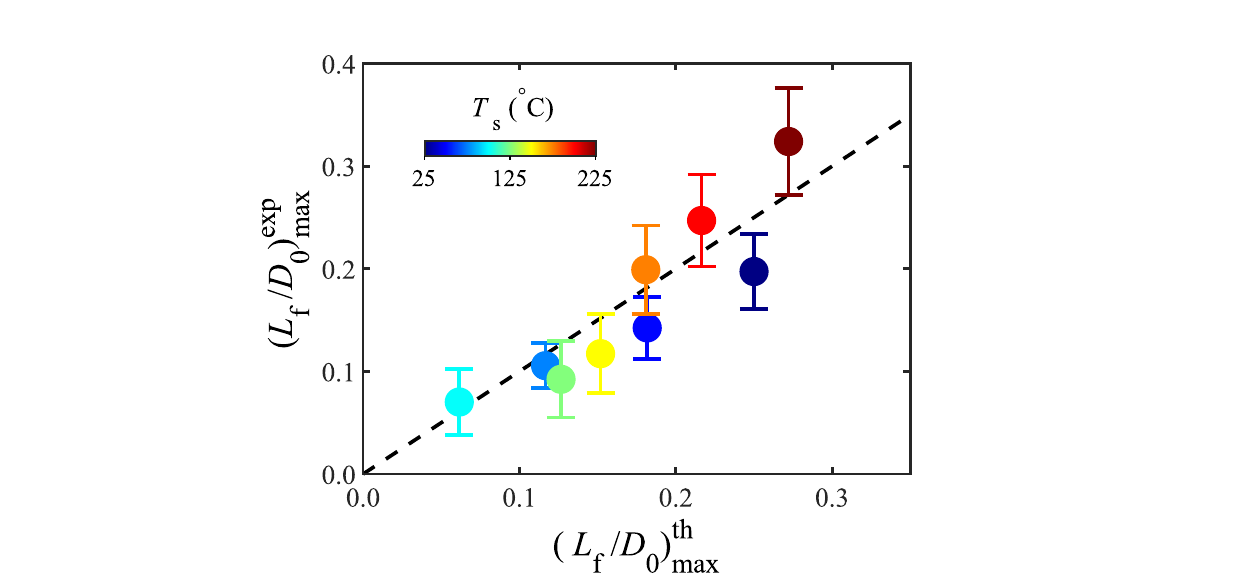}
  \caption{\textbf{Final finger length from growth speed and duration.} Measured maximum normalized finger length versus the estimate from Eq.~\eqref{eq:Lmax}. The dashed line denotes equality and the colour scale indicates substrate temperature.}
  \label{fig:4}
\end{figure*}

The nearly constant growth-termination time makes the effect of cooling transparent. In the RT-like regime, cooling slows extension and delays onset, so fingers grow more slowly for less time. In the MG regime, stronger cooling increases the estimated extension speed and advances onset, so fingers grow faster for longer. These combined effects account for the decrease and subsequent increase in final finger length. With the RT-like conversion factor calibrated as above, the model reproduces the overall magnitude and temperature dependence of the measurements as shown in Fig.~\ref{fig:4}.

This reduced-order description uses local instability scalings for an evolving rim, a mean solidification depth as a proxy for mobility contrast, and characteristic extension speeds. The solidification timing also depends on the measured spreading history. The model therefore provides approximate estimates for the present impact conditions; sensitivity to the detection threshold, solidification kinetics and solid-layer estimate is discussed in the Supplementary Information.

\section{Summary}
In summary, solidification always reduces the amount of mobile liquid, but its dynamical consequence depends on when and how the solid phase develops during impact. In the weak-undercooling regime, the solid layer remains relatively thin and smooth during the relevant instability stage, and solidification primarily reduces the effective rim deceleration and suppresses the growth of the RT-like instability. In the strong-undercooling regime, solidification develops early enough to form a structured boundary beneath the still-spreading lamella. The resulting lateral variations in hydraulic mobility redirect the residual liquid flow and promote the MG route. The unexpected re-emergence of fingering therefore reflects a transition from solidification acting predominantly as a stabilizing loss of liquid mobility to solidification reorganizing the mobility field sufficiently strongly to activate a distinct fingering pathway. More broadly, these results provide a physical basis for controlling interfacial instability and final morphology in rapidly solidifying metal droplets, with relevance to processes such as additive manufacturing like metallic 3D printing and debris mitigation in semiconducting industry.




\section*{Methods}

\textbf{Droplet impact experiment.} 
Molten tin droplets are generated by heating tin plates with the purity greater than 99.99\char37{} in a stainless steel reservoir surrounded by resistance heaters. The tin is heated above its melting temperature, $T_{\mathrm{m}} = 232~^{\circ}\mathrm{C}$. A pressure pulse generated by a solenoid valve (MHE4 series, FESTO) is used to eject the molten tin through a sapphire nozzle with an inner diameter of $400~\mu\mathrm{m}$. The pulse pressure is controlled by a mass flow controller and maintained 45--50~kPa. The pulse duration is controlled by a programmable logic controller (PLC), and is set to approximately 3~ms. These conditions can produce individual millimetre-sized droplets with an initial diameter of $D_0 = 3 \pm 0.05~\mathrm{mm}$. The release height is adjusted to obtain an impact velocity of approximately $V_0\simeq3.0\,\mathrm{m\,s^{-1}}$.

The substrate is a horizontally mounted stainless-steel sheet with dimensions of \(200~\mathrm{mm} \times 25~\mathrm{mm} \times 1~\mathrm{mm}\). Before each series of experiments, the substrate is successively polished using abrasive papers of progressively finer grit, yielding an average surface roughness $R_{\mathrm{a}} \approx 0.08~\mu\mathrm{m}$. The polished substrate is rinsed with deionized water to remove polishing debris and surface contaminants, and then dried in a nitrogen stream. To prevent residual tin and impact-induced surface modification from affecting subsequent measurements, the substrate is translated using the PLC after each impact, ensuring that every droplet can contact a previously unused region of the substrate. The temperature controlled stage is machined from a copper block with dimensions of \(100~\mathrm{mm} \times 25~\mathrm{mm} \times 8~\mathrm{mm}\). Cartridge heaters and high-sensitivity thermocouples are embedded in the copper block, and the stage temperature is regulated using a temperature controller. After repositioning to a fresh impact area, the substrate is equilibrated on the block for at least 10 min to ensure a uniform temperature before each experiment.

In addition, to suppress the formation of an oxide layer, the experiments are conducted inside a sealed chamber that is first evacuated and subsequently filled with nitrogen. The ambient pressure inside the chamber is maintained at 60--70~Pa using a vacuum pumping system and continuously monitored with a high-precision vacuum gauge. Immediately before impact, the droplets remains nearly spherical in contrast to oxidized droplets, which typically exhibits a characteristic trailing tail. This observation indicates that oxidation is effectively minimized under the present experimental conditions.

\textbf{Kinematic analysis} 
The impacting, spreading, and solidification dynamics are recorded using two high-speed cameras (FASTCAM SA1.1, Photron) operated at 6,000 fps. The cameras are equipped with the lenses (Nikon VR 105/2.8G MICRO), and backlighting is used to provide clear contrast between the droplet contour and the background. The impact process is recorded simultaneously from a direction slightly above the substrate and from a direction perpendicular to it. The side-view images are used to determine the initial droplet diameter, impact velocity, and the instant at which the droplet first contacts the substrate. The first contact between the droplet and the substrate is defined as $t=0$. The Top-view images are used to analyse the spreading contour and the temporal evolution of the droplet body, finger tips, and fingering morphology. 
The bottom morphology of the solidified droplets and the corresponding interfacial structures are further examined using an optical microscope. These images are used to characterize the smooth central region and the ridge-covered region formed during droplet spreading and solidification.

\bibliography{Reference}

\section*{Acknowledgements}

This work is supported by the Natural Science Foundation of China (NSFC) under grant nos. 12588201, 12432011, 12372219, 52450304 and 12402321. 

\section*{Author contributions}

Siyu Zhang, Feng Wang and Kai Leong Chong designed the research. 
Siyu Zhang and Hao Jiang performed the research.
Quan Zhou and Jianhua Zhang provided resources and supervised the research.
Siyu Zhang, Feng Wang and Kai Leong Chong analysed the data and wrote the manuscript.
Feng Wang, Yuan Ma and Chao Sun contributed to the interpretation of the results.
Feng Wang, Kai Leong Chong and Quan Zhou reviewed and edited the manuscript.

\begin{bibunit}[sn-mathphys-num]
\clearpage
\appendix

\section*{Supplementary information}

\setcounter{figure}{0}
\renewcommand{\thefigure}{S\arabic{figure}}

\setcounter{equation}{0}
\renewcommand{\theequation}{S\arabic{equation}}

\setcounter{section}{0}

\newcommand{\SIsection}[1]{%
    \refstepcounter{section}%
    \section*{SI \arabic{section}. #1}%
}

\SIsection{Image processing and morphology extraction}
\label{secSI1}

\begin{figure*} 
  \centering
  \includegraphics[width=1\linewidth]{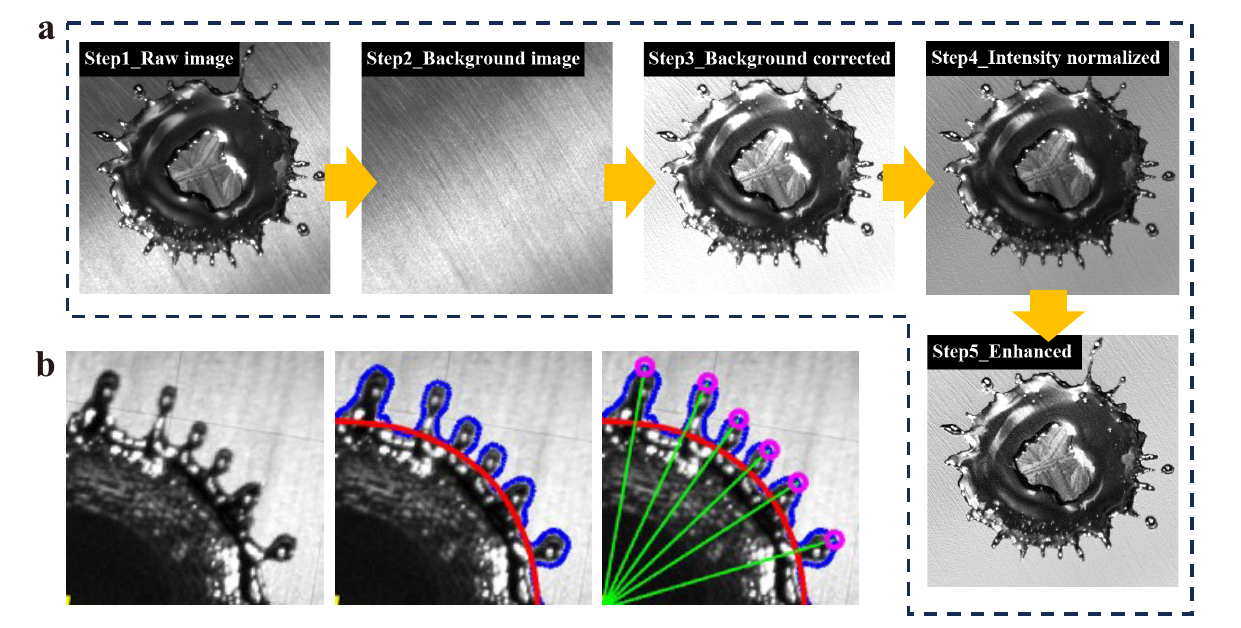}
  \caption{Image processing and extraction of droplet morphology. \textbf{a}, Sequential preprocessing of a representative high-speed image, including the raw image, background image, background correction, intensity normalization and image enhancement. \textbf{b}, Extraction of the overall droplet contour (blue), main-body contour (red), and detected finger tips (magenta). Green lines indicate the radial directions from the droplet centre to the corresponding finger tips.}
  \label{fig:S1}
\end{figure*}

The image-processing procedure used to improve the visibility of the droplet morphology is illustrated in Fig.~\ref{fig:S1}a. A raw image is first acquired under the experimental imaging conditions. For each experimental condition, a corresponding background image is recorded using the same illumination, optical configuration, camera position, and exposure settings as those used for the droplet images. The raw image is then divided pixel by pixel by the corresponding background image to obtain the background-corrected image. This correction compensates for spatial nonuniformities in illumination, optical-window transmission, and imaging response. A small numerical constant is added to the background intensity to prevent division by zero. The background-corrected images are subsequently normalized using a common intensity range defined by the 1st and 99th percentiles of the pixel intensities pooled from all processed images, yielding the intensity normalized images. Applying the same normalization limits to all frames avoids frame-by-frame contrast adjustment and ensures consistent brightness across different experimental conditions. Finally, unsharp masking is applied to enhance the local contrast at the droplet edge and peripheral fingers, producing the enhanced images. The enhanced images are used primarily for visualization. For the images presented in Fig.~\ref{fig:1}a of the main text, the brightness and contrast are additionally adjusted to improve visual consistency across the panels.

For quantitative morphology extraction, the absolute pixel-wise difference between the droplet image and the background image is first calculated and contrast-adjusted. An Otsu threshold is first determined, and \(0.7\) times this value is used to generate a binary mask to retain low-contrast regions at the tips and sides of the slender fingers. After hole filling and minor morphological correction, the largest connected component is retained as the droplet mask. The external boundary of this component is defined as the overall droplet contour, including both the main droplet body and all peripheral fingers. The contour coordinates are subsequently transformed from pixel coordinates into physical coordinates using the projective transformation obtained from the checkerboard calibration. 

The extracted overall contour is represented in polar coordinates relative to the instantaneous droplet centre. The full azimuthal range is divided into $1440$ equally spaced angular bins, corresponding to an angular resolution of $0.25^{\circ}$. For each angular direction, the maximum radial distance from the droplet centre to the overall contour is defined as $R_{\mathrm{all}}(\theta,t)$. The resulting radial profile is weakly smoothed to suppress pixel-scale fluctuations. The main-body contour is reconstructed from the lower envelope of the overall radial distribution rather than identified by a separate grayscale segmentation. For each angular position, the 20th percentile of $R_{\mathrm{all}}(\theta,t)$ within a $25^{\circ}$ moving circular window is calculated. Because the fingers produce locally large radial values, the lower-percentile envelope suppresses their influence and provides an estimate of the underlying droplet body. The resulting lower envelope is further smoothed and slightly adjusted inward to obtain a continuous representation of the main-body contour. The local reconstructed body radius is denoted by $R_{\mathrm{b}}(\theta,t)$, and the corresponding body contour is shown by the red curve in Fig.~\ref{fig:S1}b. The finger structures are identified from the radial-difference profile, $\Delta R(\theta,t)$=$R_{\mathrm{all}}(\theta,t)-R_{\mathrm{b}}(\theta,t)$. Candidate fingers are then identified as local maxima in the weakly smoothed radial-difference profile. A candidate is retained only when its radial extension is at least $0.12~\mathrm{mm}$, its peak prominence is at least $0.05~\mathrm{mm}$, and its angular separation from an adjacent candidate is at least $6^{\circ}$. These criteria prevented weak edge roughness and multiple peaks within the same broad finger from being incorrectly counted as separate fingers.

The local maxima of $\Delta R(\theta,t)$ are used to determine the approximate angular positions of the fingers. 
The precise tip position of the $n$th finger is obtained by searching for
the maximum value of $R_{\mathrm{all}}(\theta,t)$ within $\pm3^\circ$
of the corresponding detected peak. The resulting centre-to-tip distance
is denoted by $R_{t,n}(t)$.
The corresponding local main-body radius is
$R_{b,n}(t)=R_{\mathrm b}(\theta_n,t)$,
where $\theta_n$ is the angular position of the $n$th finger.
To eliminate transient edge fluctuations and isolated image artefacts, candidate fingers are tracked backward in time from a reference frame in which the fingers are fully developed, using their angular and radial continuity between successive frames. Candidates that cannot be continuously connected to a mature finger are discarded. This procedure prevents short-lived boundary irregularities from being interpreted as physical fingers. Besides, fingers exhibiting pronounced necking or incipient breakup are excluded from the calculation of the mean finger length, because their extreme elongation does not represent the typical continuous finger growth.

For each tracked finger, the radial protrusion relative to the local
reconstructed main-body contour is
$R_{t,n}(t)-R_{b,n}(t)$.
The mean finger-tip radius $R_{\mathrm t}(t)$ and mean main-body radius
$R_{\mathrm b}(t)$ are then obtained by averaging $R_{t,n}(t)$ and
$R_{b,n}(t)$, respectively, over all tracked fingers. To suppress small apparent protrusions
arising from edge fluctuations and contour reconstruction, the onset
criterion $L_{\mathrm c}=0.04D_0$ is used as the reference state. The reported
finger length is therefore defined as

\begin{equation}
L_{\mathrm f}(t)
=
R_{\mathrm t}(t)-R_{\mathrm b}(t)-L_{\mathrm c},
\qquad
t\geq t_{\mathrm{onset}}^{\mathrm{exp}},
\label{eq:S_finger_length}
\end{equation}
such that
$L_{\mathrm f}(t_{\mathrm{onset}}^{\mathrm{exp}})=0$.
The sensitivity to the choice of $L_{\mathrm c}$ is discussed in Section~\hyperref[onset_sensitivity]{SI~5}.

\SIsection{Thermophysical basis for concurrent solidification during spreading}
\label{secS21}

The impact and spreading dynamics of the molten-tin droplets evolve over a few milliseconds, with the characteristic inertial time $t_{\mathrm{i}}=D_0/V_0\simeq1~\mathrm{ms}$. To assess whether interfacial solidification can develop appreciably on this timescale, we consider the thermophysical properties governing the early growth of the solid layer.

For the one-dimensional, conduction-limited Stefan estimate, the early solid-layer thickness scales as

\begin{equation}
\delta_{\mathrm{s}}(t)
\sim
\left(
\frac{2K_{\mathrm{s}}\Delta T_{\mathrm{i}}t}
{\rho_{\mathrm{s}}L_{\mathrm{m}}}
\right)^{1/2},
\label{eq:thermophysical}
\end{equation}
where $K_{\mathrm{s}}$, $\rho_{\mathrm{s}}$, and $L_{\mathrm{m}}$ are the thermal conductivity, density, and latent heat of fusion of the solid phase, respectively, and $\Delta T_{\mathrm{i}}$ is the interfacial undercooling. This scaling shows that the early growth of the solid layer is governed by the combined effects of solid-phase heat conduction, density, and latent heat.

To place the magnitude of this material response in perspective, Table~\ref{tab:thermal_properties} compares the relevant thermophysical properties of solid tin with those of ice. Because the comparison concerns the early growth of the newly formed solidification layer, the tabulated properties refer to the solid phases rather than to the parent liquids.
\renewcommand{\tablename}{Tab.}
\renewcommand{\thetable}{S\arabic{table}}
\setcounter{table}{0}

\begin{table}[htbp]
\centering
\caption{Representative thermophysical properties relevant to early solid-layer growth in the water--ice and molten-tin systems.}
\label{tab:thermal_properties}

\small
\setlength{\tabcolsep}{8pt}
\renewcommand{\arraystretch}{1.15}

\begin{tabular}{p{5.0cm}ccc}
\hline
\textbf{Property}
&
\textbf{Symbol}
&
\textbf{Ice layer}
&
\textbf{Solid-tin layer}
\\
\hline

Solid thermal conductivity
\newline
$\left(\mathrm{W\,m^{-1}\,K^{-1}}\right)$
&
$K_{\mathrm{s}}$
&
$2.2$
&
$67$
\\

Solid density
\newline
$\left(\mathrm{kg\,m^{-3}}\right)$
&
$\rho_{\mathrm{s}}$
&
$917$
&
$7310$
\\

Solid specific heat capacity
\newline
$\left(\mathrm{J\,kg^{-1}\,K^{-1}}\right)$
&
$c_{p,\mathrm{s}}$
&
$2.05\times10^{3}$
&
$228$
\\

Solid thermal diffusivity
\newline
$\left(\mathrm{m^{2}\,s^{-1}}\right)$
&
$\alpha_{\mathrm{s}}$
&
$1.17\times10^{-6}$
&
$4.02\times10^{-5}$
\\

Latent heat of fusion
\newline
$\left(\mathrm{J\,kg^{-1}}\right)$
&
$L_{\mathrm{m}}$
&
$3.34\times10^{5}$
&
$5.9\times10^{4}$
\\

\hline
\end{tabular}

\end{table}

Solid tin has a thermal conductivity and thermal diffusivity approximately $30.5$ and $34.4$ times those of ice, respectively, whereas the latent heat of fusion of tin is only about $18\%$ of that of water. According to the Stefan scaling in Eq.~\eqref{eq:thermophysical}, the combined material factor
$\left(K_{\mathrm{s}}/\rho_{\mathrm{s}}L_{\mathrm{m}}\right)^{1/2}$
for solid tin is approximately $4.65$ times that for ice. These thermophysical properties provide favourable conditions for appreciable solid-layer growth over short timescales.

The actual solidification dynamics also depend on the interfacial undercooling and transient thermal conditions. Consistent with the Stefan estimate in the main text, an appreciable solid layer can develop within the inertial spreading time while a substantial fraction of the lamella remains liquid and mobile. Solidification can therefore evolve concurrently with radial spreading and influence the remaining liquid flow before the spreading stage terminates. 

\SIsection{Interfacial temperature and undercooling}
\label{secSI3}

The substrate temperature $T_{\mathrm{s}}$ represents the initial temperature of the substrate, whereas the interfacial temperature $T_{\mathrm{i}}$ established upon contact is determined by the initial temperatures and thermal effusivities of the droplet and substrate. Therefore, $T_{\mathrm{i}}$ or the corresponding interfacial undercooling $\Delta T_{\mathrm{i}}$, rather than $T_{\mathrm{s}}$, is used in subsequent calculations involving interfacial heat transfer, solidification, and temperature-dependent model terms. $T_{\mathrm{s}}$ is retained as the experimental control parameter and is used on the horizontal axis of the figures.

The droplet and substrate are approximated as two semi-infinite bodies, initially isothermal at \(T_{\mathrm{l}}\) and \(T_{\mathrm{s}}\), respectively, where the initial temperature of the molten tin droplet is taken as the melting temperature of tin, \(T_{\mathrm{l}}\simeq T_{\mathrm{m}}=232~^\circ\mathrm{C}\). Neglecting thermal contact resistance, the interfacial temperature is calculated as

\begin{equation}
T_{\mathrm{i}}
=
\frac{
e_{\mathrm{l}}T_{\mathrm{l}}
+
e_{\mathrm{sub}}T_{\mathrm{s}}
}{
e_{\mathrm{l}}+e_{\mathrm{sub}}
},
\end{equation}
where $e_{\mathrm{l}}$ and $e_{\mathrm{sub}}$ are the thermal effusivities of liquid tin and the substrate, respectively. The thermal effusivity is defined as

\begin{equation}
e_j
=
\left(
K_j \rho_j c_{p,j}
\right)^{1/2},
\qquad
j=\mathrm{l},\,\mathrm{sub},
\label{eq:thermal_effusivity}
\end{equation}
where $K_j$, $\rho_j$, and $c_{p,j}$ are the thermal conductivity, density, and specific heat capacity of material $j$, respectively. The corresponding interfacial undercooling is defined as

\begin{equation}
\Delta T_{\mathrm{i}}
=
T_{\mathrm{m}}-T_{\mathrm{i}}.
\label{eq:interfacial_undercooling}
\end{equation}

\setcounter{table}{0}
\renewcommand{\thetable}{S\arabic{table}}
\renewcommand{\tablename}{Tab.}

The calculated interfacial temperatures and undercoolings corresponding to the
investigated substrate temperatures are listed in
Table~\ref{tab:interface_temperature}.

\renewcommand{\tablename}{Tab.}
\renewcommand{\thetable}{S\arabic{table}}
\setcounter{table}{1}
\begin{table}[htbp]
\centering

\captionsetup{
    width=0.8\textwidth,
    labelsep=space
}
\caption{Interfacial temperature and undercooling.}
\label{tab:interface_temperature}

\begin{tabular}{c@{\hspace{1.5cm}}c@{\hspace{1.5cm}}c}
\hline
$T_{\mathrm{s}}$ ($^\circ$C) &
$T_{\mathrm{i}}$ ($^\circ$C) &
$\Delta T_{\mathrm{i}}$ (K)
\tabularnewline
\hline
25  & 122.49 & 109.51 \tabularnewline
50  & 135.71 & 96.29  \tabularnewline
75  & 148.94 & 83.06  \tabularnewline
100 & 162.16 & 69.84  \tabularnewline
125 & 175.39 & 56.61  \tabularnewline
150 & 188.62 & 43.38  \tabularnewline
175 & 201.84 & 30.16  \tabularnewline
200 & 215.07 & 16.93  \tabularnewline
225 & 228.30 & 3.70   \tabularnewline
\hline
\end{tabular}
\end{table}

\SIsection{Determination of the characteristic solidification time, $t_{\mathrm{f}}^{\mathrm{th}}$}
\label{solidification_time}

\begin{figure*} 
  \centering
  \includegraphics[width=1\linewidth]{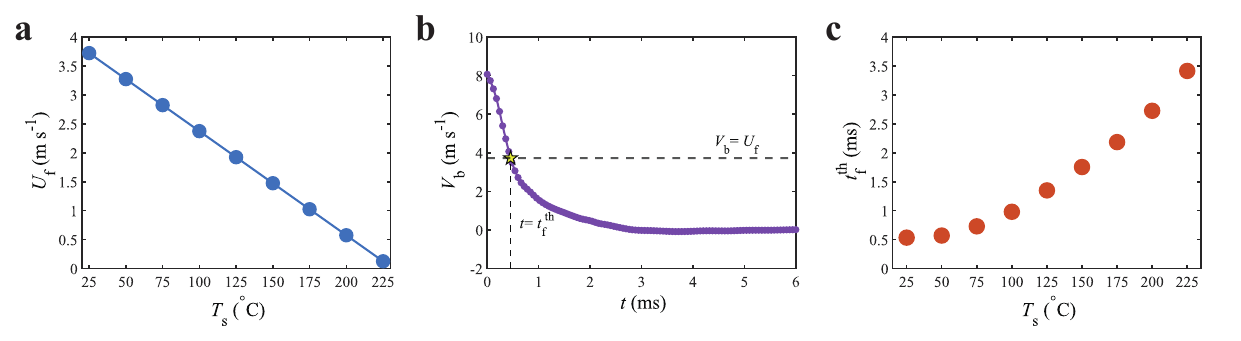}
  \caption{Determination of the characteristic solidification time. \textbf{a}, Solidification front velocity $U_{\mathrm{f}}$ as a function of substrate temperature $T_{\mathrm{s}}$, calculated using the kinetic undercooling relation $U_{\mathrm{f}}=\kappa(T_{\mathrm{m}}-T_{\mathrm{i}})$. \textbf{b}, Representative temporal evolution of the body spreading velocity $V_{\mathrm{b}}$ at $T_{\mathrm{s}}=25~ ^{\circ}\mathrm{C}$. The solidification time $t_{\mathrm{f}}^{\mathrm{th}}$ is identified as the time at which $V_{\mathrm{b}}$ decreases to the corresponding solidification front velocity $U_{\mathrm{f}}$, as indicated by the dashed lines. \textbf{c}, The resulting $t_{\mathrm{f}}^{\mathrm{th}}$ as a function of $T_{\mathrm{s}}$. The progressive increase in $t_{\mathrm{f}}^{\mathrm{th}}$ with substrate temperature reflects the reduction in interfacial undercooling and the associated decrease in the solidification front velocity.}
  \label{fig:S2}
\end{figure*}

The characteristic solidification time $t_{\mathrm{f}}^{\mathrm{th}}$ is estimated from the kinematic competition between radial spreading and solidification. For each substrate temperature, the kinetic solidification-front velocity is calculated as
\begin{equation}
U_{\mathrm{f}}
=
\kappa\Delta T_{\mathrm{i}},
\end{equation}
where $\kappa=0.034~\mathrm{m\,s^{-1}\,K^{-1}}$ is the kinetic coefficient of the solidification front \cite{gielen2020solidification}. As shown in Fig.~\ref{fig:S2}a, $U_{\mathrm{f}}$ decreases with increasing substrate temperature because the interfacial undercooling decreases.

The calculated $U_{\mathrm{f}}$ is then compared with the experimentally measured main-body spreading velocity, $V_{\mathrm{b}}(t)=\mathrm{d}R_{\mathrm{b}}/\mathrm{d}t$, at the corresponding temperature. We define $t_{\mathrm{f}}^{\mathrm{th}}$ as the time at which
\begin{equation}
V_{\mathrm{b}}
\left(
t_{\mathrm{f}}^{\mathrm{th}}
\right)
=
U_{\mathrm{f}}.
\end{equation}

Fig.~\ref{fig:S2}b illustrates this procedure for $T_{\mathrm{s}}=25\,^{\circ}\mathrm{C}$. Since $U_{\mathrm{f}}$ is constant for a given temperature condition, it appears as a horizontal line in the velocity--time plot, and its intersection with $V_{\mathrm{b}}(t)$ determines $t_{\mathrm{f}}^{\mathrm{th}}$. The same procedure is applied to all substrate temperatures.

The resulting $t_{\mathrm{f}}^{\mathrm{th}}$ increases monotonically with substrate temperature (Fig.~\ref{fig:S2}c), reflecting the reduction in interfacial undercooling and the corresponding decrease in $U_{\mathrm{f}}$. Here, $t_{\mathrm{f}}^{\mathrm{th}}$ represents the characteristic time at which solidification becomes dynamically comparable with radial spreading, rather than the first occurrence of solidification at the interface.

\SIsection{Sensitivity to the fingering-onset criterion}
\label{onset_sensitivity}

\begin{figure*} 
  \centering
  \includegraphics[width=1\linewidth]{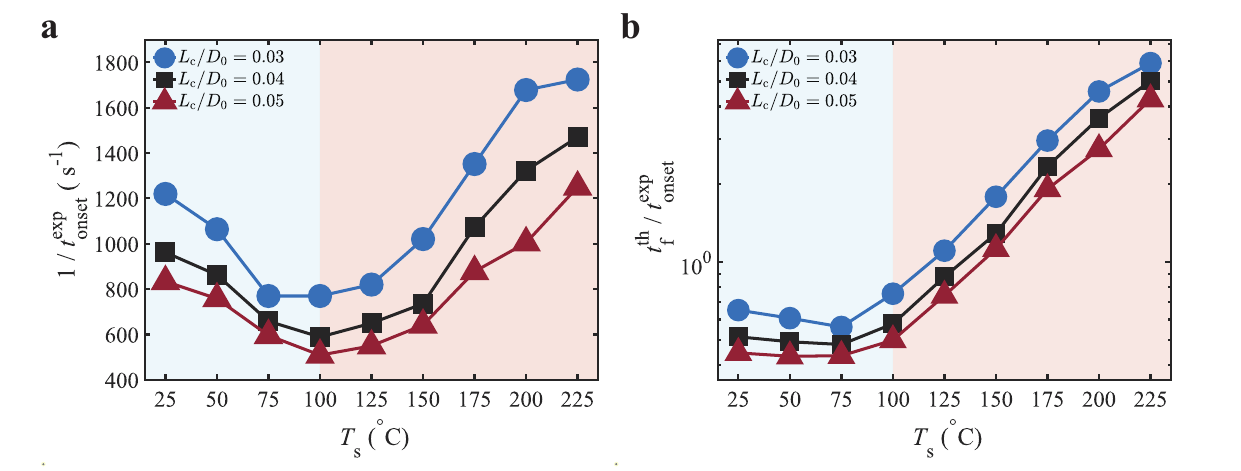}
  \caption{ Sensitivity of the fingering analysis to the onset threshold.
    \textbf{a}, Experimental onset-rate proxy,
    $1/t_{\mathrm{onset}}^{\mathrm{exp}}$.
    \textbf{b}, Solidification timing ratio,
    $t_{\mathrm{f}}^{\mathrm{th}}/t_{\mathrm{onset}}^{\mathrm{exp}}$.
    Both quantities are calculated using fingering onset thresholds of
    $L_\mathrm{c}/D_0=0.03$, $0.04$, and $0.05$.
    The blue and red shaded regions indicate the MG regime and the RT regime, respectively,
    as identified using the primary threshold $L_\mathrm{c}/D_0=0.04$.}
  \label{fig:S3}
\end{figure*}

In the experiments, the fingering-onset time is defined as the first persistent time at which the normalized finger length reaches $L_{\mathrm{c}}/D_0=0.04$. To examine whether the conclusions depend on this threshold, the analysis is repeated using $L_{\mathrm{c}}/D_0=0.03$ and $0.05$. 

We use $1/t_{\mathrm{onset}}^{\mathrm{exp}}$ as experimental onset-rate proxy. The timing ratio
$t_{\mathrm{f}}^{\mathrm{th}}/t_{\mathrm{onset}}^{\mathrm{exp}}$ is used to characterize the relative temporal ordering between solidification becoming dynamically comparable with spreading and the appearance of persistent fingering. 
Changing the onset threshold primarily shifts the values of these quantities. A larger threshold gives a later $t_{\mathrm{onset}}^{\mathrm{exp}}$ and therefore decreases both $1/t_{\mathrm{onset}}^{\mathrm{exp}}$ and $t_{\mathrm{f}}^{\mathrm{th}}/t_{\mathrm{onset}}^{\mathrm{exp}}$. 
More importantly, the temperature-dependent trends remain unchanged for all three thresholds. As shown in Fig.~\ref{fig:S3}a, $1/t_{\mathrm{onset}}^{\mathrm{exp}}$ decreases from $25\,^{\circ}\mathrm{C}$ to approximately $100\,^{\circ}\mathrm{C}$ and then increases towards $225\,^{\circ}\mathrm{C}$, with the minimum consistently located near $100\,^{\circ}\mathrm{C}$. 
Fig.~\ref{fig:S3}b also shows that $t_{\mathrm{f}}^{\mathrm{th}}/t_{\mathrm{onset}}^{\mathrm{exp}}$ exhibits the same temperature-dependent increase for all three thresholds. 
Therefore, varying the onset criterion within $L_{\mathrm{c}}/D_0=0.03$--$0.05$ changes the numerical values of $t_{\mathrm{onset}}^{\mathrm{exp}}$, $1/t_{\mathrm{onset}}^{\mathrm{exp}}$, and $t_{\mathrm{f}}^{\mathrm{th}}/t_{\mathrm{onset}}^{\mathrm{exp}}$, but does not alter the non-monotonic fingering response or the overall temporal distinction between the strong and weak-undercooling regimes. 

These results demonstrate the robustness of the observed non-monotonic fingering behaviour and the associated physical interpretation to the choice of onset threshold. The intermediate value $L_{\mathrm{c}}/D_0=0.04$ is therefore retained as the primary onset criterion, providing sufficiently early detection while avoiding the inclusion of weak edge fluctuations as persistent fingers.

\SIsection{Reduced-order model for the fingering-onset time, $t_{\mathrm{onset}}^{\mathrm{th}}$}
\label{secSI6}

The fingering-onset time, $t_{\mathrm{onset}}^{\mathrm{exp}}$, exhibits a pronounced nonmonotonic dependence on substrate temperature (Fig.~\ref{fig:3}a). In the weak-undercooling regime, $t_{\mathrm{onset}}^{\mathrm{exp}}$ decreases with increasing substrate temperature, whereas the opposite trend is observed in the strong-undercooling regime. This behaviour indicates that a single physical mechanism cannot account for the variation in onset time over the full temperature range. Instead, the experimental results suggest that the first appearance of persistent fingers is governed by two competing physical routes. At weak undercooling, fingering is governed by the Rayleigh--Taylor (RT)-like instability associated with deceleration of the spreading rim, whereas at strong undercooling, it is governed by the mobility-gradient (MG) route.

For each physical route, $t_{\mathrm{onset}}^{\mathrm{th}}$ is decomposed into the time required to initiate edge-perturbation and the subsequent amplification time needed to reach the prescribed experimental threshold. The general onset time is therefore written as

\begin{equation}
t_{\mathrm{onset}}^{\mathrm{th}}
=
t_{0,n}^{\mathrm{th}}
+
\tau_{n},
\qquad
n=\mathrm{RT},\mathrm{MG},
\label{eq:general_onset_framework}
\end{equation}
where $t_{0,n}^{\mathrm{th}}$ denotes the time at which physical route $n$ begins to drive the growth of edge perturbations, and $\tau_n$ denotes the characteristic amplification time required for the perturbation to reach the prescribed fingering criterion.
In this work, the fingering-onset criterion is
$R_{\mathrm t}-R_{\mathrm b}=L_{\mathrm c}$, with $L_{\mathrm c}=0.04D_0$, and the fingering-onset time is taken as the first persistent time at which
this criterion is satisfied.

\subsection*{RT-dominated onset, $t_{\mathrm{RT}}^{\mathrm{th}}$}
\label{secSI4_RT}

In the weak-undercooling regime, persistent fingering develops before solidification becomes sufficiently strong to reorganize the spreading flow. The dominant hydrodynamic route is therefore associated with the RT-like instability of the decelerating rim. Solidification nevertheless develops during spreading and modifies the amount of liquid that remains dynamically active. Following the general onset-time framework introduced above, the RT-dominated onset time is written as
\begin{equation}
t_{\mathrm{RT}}^{\mathrm{th}}
=
t_{0,\mathrm{RT}}^{\mathrm{th}}
+
\tau_{\mathrm{RT}}.
\label{eq:tRT_framework}
\end{equation}
For a dense liquid rim surrounded by a much lighter gas, the capillary RT dispersion relation can be written, at the scaling level, as
\begin{equation}
\sigma_{\mathrm{RT}}^{2}(k)
\sim
a_{\mathrm{eff}}k
-
\frac{\gamma}{\rho_{\mathrm{l}}}k^{3},
\label{eq:RT_dispersion}
\end{equation}
where $a_{\mathrm{eff}}$ is the magnitude of the effective rim deceleration, $\gamma$ is the surface tension, and $\rho_{\mathrm{l}}$ is the liquid density. The first term represents destabilization by rim deceleration, whereas the second term represents capillary stabilization of short-wavelength perturbations. And the corresponding growth-rate scaling
\begin{equation}
\sigma_{\mathrm{RT,max}}
\sim
\left(
\frac{\rho_{\mathrm{l}}a_{\mathrm{eff}}^{3}}
{\gamma}
\right)^{1/4}.
\label{eq:RT_sigma_m}
\end{equation}

To incorporate the effect of solidification, we approximate the spreading lamella as an immobile solid layer beneath a still-moving liquid layer. For convenience, the solidification-layer thickness evaluated at the inertial timescale is denoted by $\delta_{\mathrm{s}}(t_{\mathrm{i}})$. To quantify the reduction of the mobile-liquid layer by solidification, we use the solidification penetration ratio

\begin{equation}
\Phi_{\mathrm{i}}
=
\frac{\delta_{\mathrm{s}}(t_{\mathrm{i}})}{h_{\mathrm{i}}}
=
\left(
\frac{\mathrm{Ste}_{\mathrm{s}}}
{\mathrm{Ste}_{\mathrm{c}}}
\right)^{1/2},
\end{equation}
where $\mathrm{Ste}_{\mathrm{s}}=c_{p,\mathrm{s}}\Delta T_{\mathrm{i}}/L_{\mathrm{m}}$, and $\mathrm{Ste}_{\mathrm{c}}\simeq 1$ for the impact conditions considered here, as evaluated from the characteristic lamella thickness $h_{\mathrm{i}}$ and the inertial timescale $t_{\mathrm{i}}=D_0/V_0$.

Since the solidified layer is assumed to be immobile, whereas the remaining liquid layer retains the characteristic radial motion, the effective radial velocity can be estimated by a thickness-weighted average,
\begin{equation}
V_{\mathrm{RT}}
\sim
\frac{
\delta_{\mathrm{s}}(t_{\mathrm{i}})\times 0
+
\left(h_{\mathrm{i}}-\delta_{\mathrm{s}}(t_{\mathrm{i}})\right)V_0
}{
h_{\mathrm{i}}
}
=
V_0\left(1-\Phi_{\mathrm{i}}\right).
\end{equation}
The effective radial velocity of the mobile liquid therefore scales as
\begin{equation}
V_{\mathrm{RT}}
\sim
V_0
\left(
1-\sqrt{\mathrm{Ste}_{\mathrm{s}}}
\right).
\label{eq:VRT}
\end{equation}
Taking $R_0=D_0/2$, the characteristic deceleration in the fully liquid limit is
\begin{equation}
a_0
\sim
\frac{V_0^2}{R_0}.
\label{eq:a0_RT}
\end{equation}
The reduction in the dynamically active radial velocity gives
\begin{equation}
a_{\mathrm{eff}}
\sim
a_0
\left(
1-\sqrt{\mathrm{Ste}_{\mathrm{s}}}
\right)^2.
\label{eq:aeff_RT}
\end{equation}
Substitution into Eq.~\eqref{eq:RT_sigma_m} yields
\begin{equation}
\sigma_{\mathrm{RT,max}}
\sim
\sigma_{\mathrm{RT},0}
\left(
1-\sqrt{\mathrm{Ste}_{\mathrm{s}}}
\right)^{3/2},
\label{eq:sigma_RT_solidification}
\end{equation}
where $\sigma_{\mathrm{RT},0}$ denotes the fastest-growing rate in the fully liquid limit.

The reduction in the RT-like growth rate implies a corresponding delay in the initiation of perturbation growth. Using $R_0/V_0$ as the characteristic activation timescale in the fully liquid limit gives
\begin{equation}
t_{0,\mathrm{RT}}^{\mathrm{th}}
=
\frac{R_0/V_0}
{\left(
1-\sqrt{\mathrm{Ste}_{\mathrm{s}}}
\right)^{3/2}}.
\label{eq:t0_RT}
\end{equation}

After RT-like amplification is initiated, the perturbation must further extend to the experimental detection length. Because this extension is short compared with the overall spreading length, $V_{\mathrm{RT}}$ is taken as the characteristic radial extension speed during this stage, and its variation over this interval is neglected. The corresponding amplification time is therefore
\begin{equation}
\tau_{\mathrm{RT}}
=
\frac{L_{\mathrm{c}}}{V_{\mathrm{RT}}}
=
\frac{L_{\mathrm{c}}}
{V_0\left(
1-\sqrt{\mathrm{Ste}_{\mathrm{s}}}
\right)}.
\label{eq:tau_RT}
\end{equation}
Combining Eqs.~\eqref{eq:t0_RT} and \eqref{eq:tau_RT}, the RT-dominated fingering-onset time is
\begin{equation}
t_{\mathrm{RT}}^{\mathrm{th}}
=
\frac{R_0/V_0}
{\left(
1-\sqrt{\mathrm{Ste}_{\mathrm{s}}}
\right)^{3/2}}
+
\frac{L_{\mathrm{c}}}
{V_0\left(
1-\sqrt{\mathrm{Ste}_{\mathrm{s}}}
\right)}.
\label{eq:tRT_final}
\end{equation}

Thus, increasing solidification delays the RT-like onset through two coupled effects: it reduces the effective rim deceleration and hence the instability growth rate, while simultaneously decreasing the radial liquid velocity available for the perturbation to reach the detectable finger length.

\subsection*{MG-dominated onset, $t_{\mathrm{MG}}^{\mathrm{th}}$}
\label{secSI4_MG}

In the strong-undercooling regime, solidification develops sufficiently early to form a structured boundary beneath the still-spreading lamella. Spatial variations in the solidified-layer thickness produce variations in the thickness and hydraulic mobility of the remaining liquid layer, thereby redirecting the residual radial flow towards relatively high-mobility regions. This solidification-induced mobility heterogeneity provides the physical basis of the MG route. For the MG route, the initiation time is identified with the characteristic solidification time,
$t_{0,\mathrm{MG}}^{\mathrm{th}}=t_{\mathrm{f}}^{\mathrm{th}}$.
The MG-dominated onset time is therefore written as
\begin{equation}
t_{\mathrm{MG}}^{\mathrm{th}}
=
t_{\mathrm{f}}^{\mathrm{th}}
+
\tau_{\mathrm{MG}}.
\end{equation}

The characteristic solidification time $t_{\mathrm{f}}^{\mathrm{th}}$ and the corresponding solidification-front velocity $U_{\mathrm{f}}$ are determined in Section~\hyperref[solidification_time]{SI~4}. $t_{\mathrm{f}}^{\mathrm{th}}$ is defined as the time at which the radial spreading speed becomes comparable to the solidification-front velocity .

 We therefore use \(U_{\mathrm{f}}\) as the characteristic advection velocity in the viscous--capillary scaling for the remaining mobile-liquid layer. For a perturbation of wavenumber $k$, the temporal growth rate is written at the scaling level as
\begin{equation}
\sigma_{\mathrm{v}}(k)
\sim
U_{\mathrm{f}}k
-
\frac{\gamma h_{\mathrm{l}}^2}{\mu_{\mathrm{l}}}k^3,
\end{equation}
where $h_{\mathrm{l}}$ is the characteristic thickness of the mobile-liquid layer, $\mu_{\mathrm{l}}$ is the dynamic viscosity of liquid tin, and $\gamma$ is the surface tension. The first term represents the destabilizing contribution associated with radial advection, whereas the second term represents capillary stabilization of short-wavelength perturbations.And the fastest-growing mode gives
\begin{equation}
k_{\mathrm{m}}h_{\mathrm{l}}
\sim
\left(
\frac{\mu_{\mathrm{l}}U_{\mathrm{f}}}{\gamma}
\right)^{1/2}.
\end{equation}
So that, we can gain capillary number based on the solidification-front velocity,
\begin{equation}
\mathrm{Ca}_{\mathrm{f}}
=
\frac{\mu_{\mathrm{l}}U_{\mathrm{f}}}{\gamma},
\end{equation}
the fastest-growing wavenumber and the corresponding growth-rate scaling satisfy
\begin{equation}
k_{\mathrm{m}}h_{\mathrm{l}}
\sim
\sqrt{\mathrm{Ca}_{\mathrm{f}}},
\qquad
\sigma_{\mathrm{v,max}}
\sim
\frac{U_{\mathrm{f}}}{h_{\mathrm{l}}}
\sqrt{\mathrm{Ca}_{\mathrm{f}}}.
\end{equation}

The above relation describes the viscous--capillary amplification of the mobile layer but does not yet account explicitly for the mobility heterogeneity generated by non-uniform solidification. We use the solidification penetration ratio $\Phi_{{i}}$, defined above, as a representative proxy for the solidification-generated mobility heterogeneity. In the fully liquid limit, $\Phi_{{i}}\rightarrow0$, the solidification-induced mobility contrast would vanish. To leading order, we therefore take the mobility-gradient contribution to scale linearly with $\Phi_{{i}}$. The fastest growth rate of the MG route is then
\begin{equation}
\sigma_{\mathrm{MG,max}}
\sim
\Phi_{\mathrm{i}}\sigma_{\mathrm{v,max}}
\sim
\frac{U_{\mathrm{f}}}{h_{\mathrm{l}}}
\sqrt{\mathrm{Ca}_{\mathrm{f}}}\,
\Phi_{\mathrm{i}}.
\end{equation}
With $\mathrm{Ste}_{\mathrm{c}}\simeq1$ for the impact conditions considered here, the MG growth rate reduces to
\begin{equation}
\sigma_{\mathrm{MG,max}}
\sim
\frac{U_{\mathrm{f}}}{h_{\mathrm{l}}}
\sqrt{
\mathrm{Ca}_{\mathrm{f}}
\mathrm{Ste}_{\mathrm{s}}
}.
\end{equation}
To convert this temporal growth rate into the radial extension speed of the perturbation, we use $h_{\mathrm{l}}$ as the corresponding length scale. The MG-driven radial extension speed therefore scales as
\begin{equation}
u_{\mathrm{MG}}
\sim
h_{\mathrm{l}}
\sigma_{\mathrm{MG,max}}
\sim
U_{\mathrm{f}}
\sqrt{
\mathrm{Ca}_{\mathrm{f}}
\mathrm{Ste}_{\mathrm{s}}
}.
\end{equation}

After the MG route begins to drive the growth, the perturbation must further extend to the experimental detection length. Taking $u_{\mathrm{MG}}$ as the characteristic radial extension speed during this stage, and neglecting its variation, the corresponding extension time is

\begin{equation}
\tau_{\mathrm{MG}}
=
\frac{L_{\mathrm{c}}}{u_{\mathrm{MG}}}
=
\frac{L_{\mathrm{c}}}
{
U_{\mathrm{f}}
\sqrt{
\mathrm{Ca}_{\mathrm{f}}
\mathrm{Ste}_{\mathrm{s}}
}
}.
\end{equation}
Combining $t_{\mathrm{f}}^{\mathrm{th}}$ and $\tau_{\mathrm{MG}}$, the MG-dominated fingering-onset time is therefore
\begin{equation}
t_{\mathrm{MG}}^{\mathrm{th}}
=
t_{\mathrm{f}}^{\mathrm{th}}
+
\frac{L_{\mathrm{c}}}
{
U_{\mathrm{f}}
\sqrt{
\mathrm{Ca}_{\mathrm{f}}
\mathrm{Ste}_{\mathrm{s}}
}
}.
\end{equation}
Thus, increasing interfacial undercooling advances the MG-dominated onset through two coupled effects: it causes solidification to become dynamically comparable with spreading at an earlier stage, while simultaneously enhancing the MG-driven growth of the perturbation.

\begin{figure*} 
  \centering
  \includegraphics[width=1\linewidth]{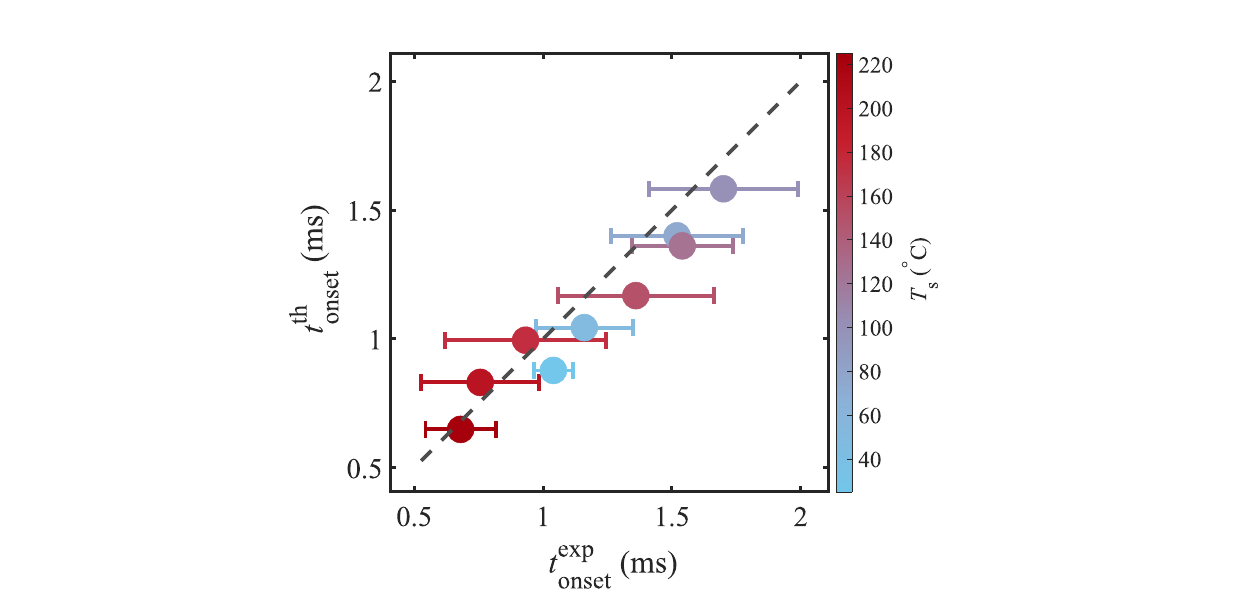}
  \caption{ Validation of the reduced-order model for fingering-onset time. Predicted versus experimentally measured fingering-onset times over the investigated substrate-temperature range. The dashed line denotes equality, and the colour scale indicates the substrate temperature. Error bars represent the experimental variability.}
  \label{fig:S4}
\end{figure*}

Thus, the predicted fingering-onset time is taken as,

\begin{equation}
t_{\mathrm{onset}}^{\mathrm{th}}
=
\begin{cases}
t_{\mathrm{RT}}^{\mathrm{th}}
=
\dfrac{R_0/V_0}
{\left(1-\sqrt{\mathrm{Ste}_{\mathrm{s}}}\right)^{3/2}}
+
\dfrac{L_{\mathrm{c}}}
{V_0\left(1-\sqrt{\mathrm{Ste}_{\mathrm{s}}}\right)},
& \text{RT-like route},
\\[12pt]
t_{\mathrm{MG}}^{\mathrm{th}}
=
t_{\mathrm{f}}^{\mathrm{th}}
+
\dfrac{L_{\mathrm{c}}}
{U_{\mathrm{f}}\sqrt{Ca_{\mathrm{f}}\mathrm{Ste}_{\mathrm{s}}}},
& \text{MG route}.
\label{eq:tonset}
\end{cases}
\end{equation}
To evaluate the predictive capability of the model, the calculated onset times from Eq.~\eqref{eq:tonset} are compared with the experimental measurements. As shown in Fig.~\ref{fig:3}a, the model reproduces the non-monotonic variation of $t_{\mathrm{onset}}^{\mathrm{th}}$ with substrate temperature. Fig.~\ref{fig:S4} further shows that the predicted onset times agree well with the measurements over the investigated temperature range. These results demonstrate that the reduced-order model captures the variation in the fingering-onset time, and supports the proposed RT-like and MG routes.

\SIsection{Calculation of the maximum finger length}
\label{max_finger_length_model}

The onset-time model developed in the preceding section determines the physical route that first reaches the prescribed fingering criterion and the corresponding theoretical onset time. Based on this result, we further estimate the maximum finger length by describing the post-onset extension of the same RT-like or MG route until finger growth terminates. We therefore construct the
maximum-length model directly from the post-onset kinematics of the
finger.

The experimental fingering onset is defined as the first persistent time at which the finger length reaches the prescribed threshold $L_{\mathrm{c}}$. The theoretical onset model adopts the same finite-length criterion, such that

\begin{equation}
L_{\mathrm{f}}
\left(
t_{\mathrm{onset}}^{\mathrm{th}}
\right)
=
0.
\label{eq:S7_onset_length}
\end{equation}
Accordingly, the maximum-length model starts from $L_{\mathrm{c}}$ and describes only the additional finger extension after onset.

After the onset criterion is reached, the finger continues to extend
until the growth-termination time $t_{\mathrm{end}}^{\mathrm{th}}$.
The maximum finger length can therefore be written as

\begin{equation}
L_{\mathrm{f},\max}^{\mathrm{th}}
=
\int_{t_{\mathrm{onset}}^{\mathrm{th}}}
^{t_{\mathrm{end}}^{\mathrm{th}}}
u_{\mathrm{sel}}(t)\,
\mathrm{d}t,
\label{eq:S7_growth_integral}
\end{equation}
where $u_{\mathrm{sel}}(t)$ denotes the radial extension velocity
associated with the fingering route selected by the onset-time model.
Using its characteristic post-onset value $u_{\mathrm{sel}}$, the
maximum normalized finger length is written as

\begin{equation}
\left(
\frac{L_{\mathrm{f}}}{D_0}
\right)_{\max}^{\mathrm{th}}
=
\frac{u_{\mathrm{sel}}}{D_0}
\left(
t_{\mathrm{end}}^{\mathrm{th}}
-
t_{\mathrm{onset}}^{\mathrm{th}}
\right).
\label{eq:S7_Lmax_general}
\end{equation}

The theoretical fingering-onset time $t_{\mathrm{onset}}^{\mathrm{th}}$ used above is given by Eq.~\eqref{eq:tonset}, corresponding to the route that reaches the detection threshold first. Accordingly, the characteristic velocity $u_{\mathrm{sel}}$ in
Eq.~\eqref{eq:S7_Lmax_general} is taken from the same route that
determines $t_{\mathrm{onset}}^{\mathrm{th}}$. The RT-like and MG
routes are considered separately below.

The time at which finger extension terminates varies only weakly with
substrate temperature compared with the pronounced variation of the
fingering-onset time. We therefore describe the growth-termination
time using the inertial--capillary timescale

\begin{equation}
\tau_{\gamma}
=
\left(
\frac{\rho_{\mathrm{l}}D_0^3}{\gamma}
\right)^{1/2},
\label{eq:S7_capillary_time}
\end{equation}
and take
\begin{equation}
t_{\mathrm{end}}^{\mathrm{th}}
=
0.20\tau_{\gamma}
=
0.20
\left(
\frac{\rho_{\mathrm{l}}D_0^3}{\gamma}
\right)^{1/2}.
\label{eq:S7_tend}
\end{equation}

Experimentally, $t_{\mathrm{end}}^{\mathrm{exp}}$ varies only weakly with substrate temperature, remaining within approximately $3.5\text{--}4.0~\mathrm{ms}$. For the present droplet properties, $t_{\mathrm{end}}^{\mathrm{th}}\simeq3.74~\mathrm{ms}$.The same growth-termination time is used for both the RT-like and MG
routes, so that their different predictions for the maximum finger
length arise from their different onset times and characteristic
post-onset extension velocities.

\subsection*{Maximum finger length in the RT-like route}
\label{secSI7_RT}

We first consider the conditions for which the RT-like route determines the fingering onset. According to the onset model developed in Section~\hyperref[secSI6]{SI~6}, the corresponding theoretical onset time is

\begin{equation}
t_{\mathrm{onset}}^{\mathrm{th}}
=
t_{\mathrm{RT}}^{\mathrm{th}}
=
\frac{
R_0/V_0
}{
\left(
1-\sqrt{\mathrm{Ste}_{\mathrm{s}}}
\right)^{3/2}
}
+
\frac{
L_{\mathrm{c}}
}{
V_0
\left(
1-\sqrt{\mathrm{Ste}_{\mathrm{s}}}
\right)
}.
\label{eq:S7_RT_selection}
\end{equation}

As derived in the preceding section, the characteristic radial velocity of the dynamically mobile liquid in the RT-like route scales as
$V_{\mathrm{RT}}\sim V_0\left(1-\sqrt{\mathrm{Ste}_{\mathrm{s}}}\right)$.
However, $V_{\mathrm{RT}}$ represents the radial transport of the
remaining mobile liquid, whereas $R_{\mathrm t}-R_{\mathrm b}$
represents the geometric radial protrusion and the reported finger
length $L_{\mathrm f}$ measures its post-onset growth relative to the
onset threshold. Therefore, only the differential motion between the
finger tip and the main droplet body contributes to the increase in
$L_{\mathrm f}$.
We therefore relate the characteristic post-onset finger-extension velocity to $V_{\mathrm{RT}}$ through

\begin{equation}
u_{\mathrm{RT}}
=
\eta_{\mathrm{RT}}V_{\mathrm{RT}},
\label{eq:S7_RT_eta}
\end{equation}
where $\eta_{\mathrm{RT}}$ is the kinematic conversion factor between the radial transport of the mobile liquid and the relative extension of the finger.

The conversion factor $\eta_{\mathrm{RT}}$ is determined from the
time-resolved finger-length measurements. For each condition in the
RT-like branch, the mean post-onset finger-extension velocity is
evaluated as

\begin{equation}
\overline{u}_{\mathrm{RT},\exp}
=
\frac{
L_{\mathrm{f},\max}^{\exp}
}{
t_{\mathrm{end}}^{\exp}
-
t_{\mathrm{onset}}^{\exp}
}.
\label{eq:S7_RT_uexp}
\end{equation}
The measured $\overline{u}_{\mathrm{RT},\exp}$ is approximately
proportional to $V_{\mathrm{RT}}$, and a least-squares fit constrained
through the origin gives $\eta_{\mathrm{RT}}=0.103\simeq0.10$.
Accordingly, using
$V_{\mathrm{RT}}\sim
V_0(1-\sqrt{\mathrm{Ste}_{\mathrm{s}}})$,
the characteristic post-onset extension velocity in the RT-like route
is

\begin{equation}
u_{\mathrm{RT}}
\sim
\eta_{\mathrm{RT}}
V_0
\left(
1-\sqrt{\mathrm{Ste}_{\mathrm{s}}}
\right).
\label{eq:S7_RT_usel}
\end{equation}
The maximum normalized finger length in the RT-like route is therefore given by

\begin{equation}
\left(
\frac{L_{\mathrm{f}}}{D_0}
\right)_{\mathrm{RT},\max}^{\mathrm{th}}
=
\eta_{\mathrm{RT}}
\frac{
V_0
\left(
1-\sqrt{\mathrm{Ste}_{\mathrm{s}}}
\right)
}{D_0}
\left(
t_{\mathrm{end}}^{\mathrm{th}}
-
t_{\mathrm{RT}}^{\mathrm{th}}
\right).
\label{eq:S7_RT_Lmax}
\end{equation}

Eq.~\eqref{eq:S7_RT_Lmax} shows that the maximum finger length in the RT-like route is controlled by both the characteristic post-onset extension velocity and the remaining growth interval. Increasing solidification decreases $u_{\mathrm{sel}}$ by reducing the radial transport of the mobile liquid, while shortening the interval $t_{\mathrm{end}}^{\mathrm{th}}-t_{\mathrm{RT}}^{\mathrm{th}}$. These two effects act together to reduce the maximum finger extension in the RT-like route.

\subsection*{Maximum finger length in the MG route}
\label{secSI7_RT}

For the MG route, the theoretical fingering-onset time is given by,

\begin{equation}
t_{\mathrm{onset}}^{\mathrm{th}}
=
t_{\mathrm{MG}}^{\mathrm{th}}
=
t_{\mathrm{f}}^{\mathrm{th}}
+
\frac{
L_{\mathrm{c}}
}{
U_{\mathrm{f}}
\sqrt{
\mathrm{Ca}_{\mathrm{f}}
\mathrm{Ste}_{\mathrm{s}}
}
}.
\label{eq:S7_MG_selection}
\end{equation}
The characteristic radial extension velocity obtained from the MG growth scaling is
$u_{\mathrm{MG}}\sim
U_{\mathrm f}
\sqrt{\mathrm{Ca}_{\mathrm{f}}\mathrm{Ste}_{\mathrm{s}}}$.
Unlike $V_{\mathrm{RT}}$, which represents the radial transport velocity of the remaining mobile liquid, $u_{\mathrm{MG}}$ is already defined as the characteristic radial extension velocity of the perturbation. Therefore, no additional kinematic conversion factor is required.

The maximum normalized finger length in the MG route is therefore given by

\begin{equation}
\left(
\frac{L_{\mathrm{f}}}{D_0}
\right)_{\mathrm{MG},\max}^{\mathrm{th}}
=
\frac{
U_{\mathrm{f}}
\sqrt{
\mathrm{Ca}_{\mathrm{f}}
\mathrm{Ste}_{\mathrm{s}}
}
}{D_0}
\left(
t_{\mathrm{end}}^{\mathrm{th}}
-
t_{\mathrm{MG}}^{\mathrm{th}}
\right).
\label{eq:S7_MG_Lmax}
\end{equation}
Eq.~\eqref{eq:S7_MG_Lmax} shows that the maximum finger length in the MG route is controlled by both the characteristic post-onset extension velocity and the remaining growth interval. Increasing interfacial undercooling increases $U_{\mathrm{f}}$, $\mathrm{Ca}_{\mathrm{f}}$, and $\mathrm{Ste}_{\mathrm{s}}$, thereby increasing $u_{\mathrm{sel}}$, while simultaneously advancing $t_{\mathrm{MG}}^{\mathrm{th}}$ and extending the interval $t_{\mathrm{end}}^{\mathrm{th}}-t_{\mathrm{MG}}^{\mathrm{th}}$. These two effects act together to increase the maximum finger extension in the MG route.

\begin{equation}
\begin{aligned}
\left(
\frac{L_\mathrm f}{D_0}
\right)_{\max}^{\mathrm{th}}
&=
\frac{u_{\mathrm{sel}}}{D_0}
\left(
t_{\mathrm{end}}^{\mathrm{th}}
-
t_{\mathrm{onset}}^{\mathrm{th}}
\right),
\\[10pt]
t_{\mathrm{onset}}^{\mathrm{th}}
&=
\begin{cases}
t_{\mathrm{RT}}^{\mathrm{th}},
& \text{RT-like route},
\\[6pt]
t_{\mathrm{MG}}^{\mathrm{th}},
& \text{MG route},
\end{cases}
\\[10pt]
u_{\mathrm{sel}}
&=
\begin{cases}
\eta_{\mathrm{RT}}V_0
\left(
1-\sqrt{\mathrm{Ste}_{\mathrm{s}}}
\right),
& \text{RT-like route},
\\[6pt]
U_{\mathrm{f}}
\sqrt{
{\mathrm{Ca}}_{\mathrm{f}}\mathrm{Ste}_{\mathrm{s}}
},
& \text{MG route}.
\end{cases}
\end{aligned}
\label{eq:S7_Lmax_unified}
\end{equation}

Eq.~\eqref{eq:S7_Lmax_unified} shows that the maximum finger length is determined by the post-onset extension velocity of the selected fingering route and the time remaining between fingering onset and growth termination. In the RT-like route, increasing solidification reduces the radial transport of the mobile liquid and delays fingering onset, so that both the characteristic extension velocity and the available growth interval decrease. In contrast, in the MG route, increasing interfacial undercooling enhances the characteristic perturbation-extension velocity and advances fingering onset, thereby increasing both contributions to the final finger length. The opposite responses of the two routes therefore account for the non-monotonic variation of the maximum finger length with substrate temperature.

\putbib[Reference_S]
\end{bibunit}

\end{document}